\documentclass[11pt, a4paper]{article}

\usepackage{jheppub}
\author[a]{Marc-Antoine Fiset,}
\author[b]{Mateo Galdeano}
\affiliation[a]{Institut f\"{u}r Theoretische Physik, ETH Z\"{u}rich,\\
Wolfgang-Pauli-Stra{\ss}e 27, 8093 Z\"{u}rich, Switzerland}
\affiliation[b]{Mathematical Institute, Oxford University\\Andrew Wiles Building, Woodstock Road\\Oxford OX2 6GG, UK}
\emailAdd{mfiset@phys.ethz.ch}
\emailAdd{mateo.galdeano@maths.ox.ac.uk}

\usepackage{setspace} 
\usepackage[normalem]{ulem} 
\usepackage{amsmath} 
\usepackage{nccmath} 
\usepackage{amssymb} 
\usepackage{amsthm} 
\usepackage{upgreek} 
\usepackage{datetime} \settimeformat{ampmtime}

\usepackage[new]{old-arrows} 
\usepackage{dsfont} 
\usepackage{bm} 
\usepackage{pbox} 
\usepackage{tikz} 
\usetikzlibrary{arrows} 
\usepackage{colortbl} 
\usepackage{mathrsfs} 
\usepackage{mathtools} 

\usepackage[utf8]{inputenc} 
\usepackage{fancyhdr} 
\usepackage{changepage} 

\newcommand{\tr}[0]{\text{tr}} 
\newcommand{\dd}{{\rm d}} 
\DeclarePairedDelimiterX\braket[2]{\langle}{\rangle}{#1 \delimsize\vert #2} 
\newcommand{\SVseven}{\text{SV}^{\text{G}_2}}
\newcommand{\SVeight}{\text{SV}^\text{Spin(7)}}

\usepackage{mathrsfs}
\usepackage{stmaryrd}
\usepackage{cleveref}

\newcommand{\normord}[1]{:\mathrel{#1}:} 

\newcommand*\Sc{\text{S}}
\newcommand*\R{\mathbb{R}}

\begin{document}

\title{Superconformal algebras for generalized Spin(7) and G$_2$ connected sums}

\abstract{Worldsheet string theory compactified on exceptional holomony manifolds is revisited following \cite{Fiset:2018huv}, where aspects of the chiral symmetry were described for the case where the compact space is a 7-dimensional G$_2$-holonomy manifold constructed as a Twisted Connected Sum. We reinterpret this result and extend it to Extra Twisted Connected Sum G$_2$-manifolds, and to 8-dimensional Generalized Connected Sum Spin(7)-manifolds. Automorphisms of the latter construction lead us to conjecture new mirror maps.
}

\maketitle

\section{Introduction}

Important steps {have been} taken in recent years to understand string theory implications of the mathematical construction \cite{MR2024648, MR3109862, Corti:2012kd} of new 7-dimensional closed manifolds with G$_2$ holonomy via the ``Twisted Connected Sum'' (TCS) approach and its generalizations.\footnote{A non-exhaustive sample of these interesting works is  \cite{Halverson:2014tya, Halverson:2015vta, Braun:2016igl, Braun:2017ryx, Braun:2017uku, Guio:2017zfn, Braun:2017csz, Braun:2018fdp, Braun:2018joh, Acharya:2018nbo, Braun:2018vhk, Braun:2019lnn, Barbosa:2019bgh, Braun:2019wnj, Xu:2020nlh, Hubner:2020yde, Cvetic:2020piw}.
} Of particular relevance to this paper are the worldsheet aspects of type II backgrounds of the form
\begin{equation}
\mathbb{M}_{10-d} \times \mathcal{M}_d \ ,
\end{equation}
where $\mathbb{M}_{10-d}$ denotes Minkowski space and $\mathcal{M}_d$ is $d$-dimensional and has exceptional holonomy. Worldsheet considerations of such backgrounds started with the work of Shatashvili and Vafa \cite{Shatashvili:1994zw}, see also \cite{Figueroa-OFarrill:1996tnk, Blumenhagen:1991nm, Figueroa-OFarrill:1990tqt, Figueroa-OFarrill:1990mzn}, who identified the chiral symmetry W-algebras characteristic of $\mathcal{M}_d$ having holonomy either G$_2$ ($d=7$) or Spin(7) ($d=8$) but being otherwise generic. These algebras are two specific extensions of the chiral $\mathcal{N}=1$ superconformal symmetry of the $\sigma$-model with target space $\mathcal{M}_d$, {and we shall denote them} by $\SVseven$ and $\SVeight$. Subsequent research focused on various aspects of theories with these symmetries \cite{Partouche:2000uq, Gepner:2001px, Eguchi:2001xa, Sugiyama:2001qh, Noyvert:2002mc, Sugiyama:2002ag, Eguchi:2003yy, deBoer:2005pt, deBoer:2006bp, Sriharsha:2006zc, Benjamin:2014kna, Cheng:2015fha}, particularly orbifolds of free CFTs by a finite group \cite{Shatashvili:1994zw, Acharya:1996fx, Acharya:1997rh, Gaberdiel:2004vx, Chuang:2004th}, and also realizing $\mathcal{M}_7=\big(\text{S}^1\times (\text{Gepner model})\big)/\mathbb{Z}_2$ \cite{Roiban:2001cp, Eguchi:2001ip, Blumenhagen:2001jb, Roiban:2002iv} or $\mathcal{M}_8=\big(\text{Gepner model}\big)/\mathbb{Z}_2$ \cite{Blumenhagen:2001qx}. Both of these reflect geometric constructions pioneered by Joyce \cite{MR1424428, MR1383960, Joyce2007}. It is natural to then also consider the CFT emerging when $\mathcal{M}_d$ is a connected sum; this was initiated in \cite{Braun:2017ryx, Braun:2017csz, Fiset:2018huv, Braun:2019lnn}. A particularly intriguing application of these ventures was to test the G$_2$ and Spin(7) analogues of mirror symmetry conjectured in \cite{Shatashvili:1994zw, Papadopoulos:1995da}.

\begin{figure}[h]
\begin{center}
\begin{tikzpicture}
\draw [black]
(2,0.5) to (1,0.5)
to[out=180, in=-10] (-0.5,0.75)
to[out=170, in=90] (-1.5,0)
to[out=-90, in=190] (-0.5,-0.75)
to[out=10, in=180] (1,-0.5) to (2,-0.5);
\draw [black]
(0.75,0.4) to (1.75,0.4)
to[out=0, in=190] (3.25,0.65)
to[out=10, in=90] (4.25,-0.1)
to[out=-90, in=-10] (3.25,-0.85)
to[out=170, in=0] (1.75,-0.6) to (0.75,-0.6);
\node at (1.375,0.8) {$\overbrace{\qquad\qquad\qquad}$};
\node at (1.375,-1.0) {$\underbrace{\qquad\qquad\qquad\qquad\qquad\qquad\qquad}$};
\node at (1.375,-2.5) {\textbf{(a)}};
\node at (10,-2.5) {\textbf{(b)}};
\node at (9,0.6) {\rotatebox{45}{$\subset$}};
\node at (11,0.6) {\rotatebox{135}{$\subset$}};
\node at (9,-0.7) {\rotatebox{135}{$\subset$}};
\node at (11,-0.7) {\rotatebox{45}{$\subset$}};

\node at (1.375,1.25) {$\text{CY}_2\times \mathbb{T}^2\times \mathbb{R}$};
\node at (-0.4,0) {$(\text{CY}_3 \times \text{S}^1)_+$};
\node at (3.1,-0.1) {$(\text{CY}_3 \times \text{S}^1)_-$};
\node at (1.375,-1.5) {G$_2$ TCS};

\node at (10,1.25) {$\text{Od}_2\oplus \text{Fr}^3$};
\node at (8,-0.05) {$(\text{Od}_3\oplus \text{Fr}^1)_+$};
\node at (12,-0.05) {$(\text{Od}_3\oplus \text{Fr}^1)_-$};
\node at (10,-1.4) {$\SVseven$};
\end{tikzpicture}
\caption{\textbf{(a)} Sketch of a compact 7-dimensional G$_2$-holonomy manifold obtained as Twisted Connected Sum (TCS). \textbf{(b)} Diamond of algebra inclusions corresponding to a $\sigma$-model whose target space is a TCS.}
\label{fig:TCS}
\end{center}
\end{figure}
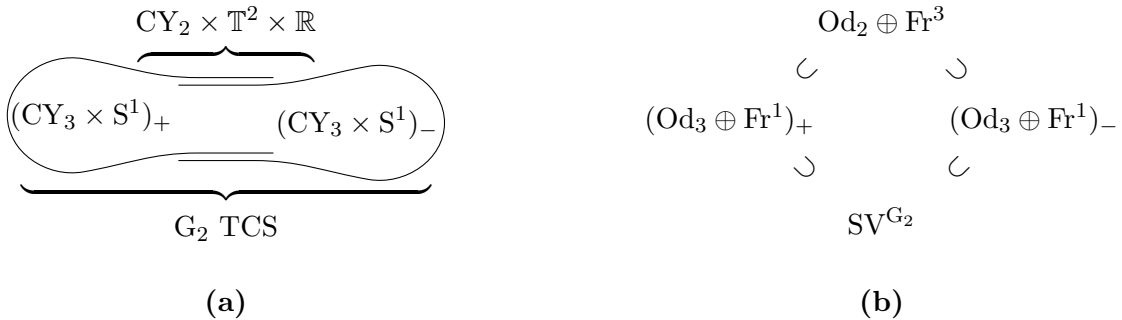

In this work we extend upon \cite{Fiset:2018huv}, where general symmetry aspects of the $\sigma$-model in a generic TCS  manifold were first investigated. Very superficially a G$_2$ TCS has the structure sketched in Figure~\ref{fig:TCS}(a): two open manifolds $\mathcal{M}_\pm = (\text{CY}_3 \times \text{S}^1)_\pm$ glued together by an appropriate isomorphism along a ``neck'' region where they have asymptotically the form of a cylinder with cross-section $\text{CY}_2\times \mathbb{T}^2$. (By $\text{CY}_n$ we mean Calabi-Yau manifold of complex dimension $n$).

\noindent These local geometric models were argued to translate to specific W-algebras on the worldsheet, in which $\SVseven$ was found to sit as a subalgebra. The transition map between the two halves of the geometry were said to correspond to automorphisms of the various chiral algebras. While this is essentially correct we find it useful to revisit that interpretation in Section~\ref{sec:Idea}, stressing the uniqueness of the $\SVseven$ subalgebra arising in these considerations. We explain that the worldsheet theory displays, at least in the appropriate regime, a network of algebra inclusions in the shape of a diamond, see Figure~\ref{fig:TCS}(b). The upper tip of the diamond is the symmetry of the $\sigma$-model excitations localized on the neck of the TCS, whereas the left and right lateral tips reflect the symmetry of the two open subsets $(\text{CY}_3\times\text{S}^1)_\pm$. In the intersection of the latter two subalgebras, a unique $\SVseven$ sits, reflecting symmetries enjoyed by the $\sigma$-model whose target space is the whole compact $\mathcal{M}_7$.

We also generalize this set-up to the case where $\mathcal{M}_7$ is an ``Extra Twisted Connected Sum'' (ETCS) \cite{Crowley:2015ctv, Nordstrom:2018cli, goette2020nuinvariants}. This is a worthy addition given that certain topological types of G$_2$-manifolds are known to admit the ETCS construction, but not the TCS construction. We briefly describe the ETCS construction in Section~\ref{sec:G2}, concentrating on differences with the ordinary TCS case which are perceptible in CFT language. We then find that the diamond of algebra inclusions is unaffected by these changes, thus asserting its general validity.

Our other focus is on Spin(7)-holonomy manifolds $\mathcal{M}_8$. In that case a conjectural construction called ``Generalized Connected Sum'' (GCS) was proposed by Braun and Sch\"afer-Nameki, mimicking the TCS construction \cite{Braun:2018joh}. We review it in Section~\ref{sec:Spin7Geom}; it is essentially like in Figure~\ref{fig:TCS}(a), except that the local models are different, see Figure~\ref{fig:GCS}(a). While various arguments were provided in \cite{Braun:2018joh}, no mathematical proof currently shows that this systematically yields Spin(7)-holonomy manifolds. We provide here what can be regarded as proof from worldsheet string theory. We describe the diamond of chiral algebra inclusions it yields in Section~\ref{sec:Spin7Alg}. Similarly to the G$_2$ case we find that a unique $\SVeight$ features in an appropriate intersection of subalgebras, at the bottom of the diamond. In the rest of Section~\ref{sec:Spin7} we examine various algebra automorphisms and interpret them with a view towards Spin(7) mirror symmetry.
Studying four examples of Joyce orbifolds which admit a GCS description leads us to propose three new mirror maps, as well as recovering some aspects of \cite{Braun:2019lnn} where similar questions were tackled.

Finally in Section~\ref{sec:Num} we study numerically whether the Shatashvili-Vafa algebra at the bottom tip of all the diamonds we have mentioned so far, which sits in the intersection between the algebras on the lateral tips, is in fact equal to that intersection. This indeed seems to be the case, at least up to level 5 in the vaccum module in the TCS case; and up to level 6 in the GCS case. If true, this would mean that TCS G$_2$-manifolds are representative of generic G$_2$-manifolds, and that GCS Spin(7)-manifolds are representative of generic Spin(7)-manifolds, at least as far as chiral symmetries of strings into them are concerned.

\subsection{The general idea} \label{sec:Idea}

In this section we set up notations and the dictionary between target space and worldsheet symmetries which will be useful throughout the rest of the paper.

The general philosophy is in fact not specific to two dimensions; consider a quantum field theory whose fields are valued, by definition, in configuration space $\mathcal{M}$. Then the ``simpler'' $\mathcal{M}$ is, the more symmetric the field theory will be. Various invariances can be counted in the set of ``symmetries'', but for us these are the chiral algebra, say on the holomorphic side, of an $\mathcal{N}=(1,1)$ superconformal $\sigma$-model with Riemannian target space $\mathcal{M}$. Let us denote them by
\begin{equation}
\Rbag(\mathcal{M}) \ .
\end{equation}
Similarly ``simplicity'' can take various meanings:
for us it refers to a reduction of the G-structure of $\mathcal{M}$ from $\text{G}=\text{O}(d)$,\footnote{O($d$) reflects the Riemannian metric $g$ on $\mathcal{M}$.} to a subgroup. Equivalently,\footnote{See for instance \cite{Joyce2007}.} there is a tangent bundle connection on $\mathcal{M}$ with holonomy strictly contained in $\text{O}(d)$. Covariantly constant differential $p$-forms under that connection then lead to worldsheet symmetries \cite{Howe:1991ic}.\footnote{This principle was recently shown \cite{delaOssa:2018azc} to hold even in the most general $\mathcal{N}=(1,0)$ non-linear $\sigma$-model, and at 1-loop in $\sqrt{\alpha'}$ perturbation theory.} 
The worldsheet currents come in supersymmetric pairs, and have spin $\frac{p}{2}$ and $\frac{p+1}{2}$. The basic holomorphic $\mathcal{N}=1$ superconformal symmetry then gets replaced by some other $\mathcal{N}=1$ superconformal W-algebra.

The simplest example is when $\mathcal{M} = \mathbb{R}\times \mathcal{X}$ is a cylinder over an arbitrary manifold $\mathcal{X}$.
The 1-form $\dd t$ along $\mathbb{R}$ is constant and thus leads to worldsheet currents
\begin{equation}
\psi_t \qquad \text{and}\qquad
j_t = i\partial t \ ,
\end{equation}
which are a free Majorana-Weyl fermion and a $\widehat{\mathfrak{u}}(1)$ current. Together they generate the chiral algebra that we shall denote $\text{Fr}^1$; in general
\begin{equation}
\text{Fr}^n = \underbrace{\big(\text{free fermion} \oplus \widehat{\mathfrak{u}}(1)\big)\oplus\ldots}_{n \text{ times}} \ .
\end{equation}
The superconformal symmetry is given by the standard expressions\footnote{We reuse these notations throughout the paper. If $t$ parametrizes $\mathbb{R}$ or $\text{S}^1$ in target space, $G_t$ is the chiral supersymmetry current {and} $T_t$ is the holomorphic part of the stress-tensor. Colons represent normal ordering and $\partial$ is the derivative with respect to the holomorphic worldsheet coordinate.}
\begin{equation} \label{eq:virasorostandard}
G_t = \normord{j_t\psi_t} \ , \qquad
T_t = \frac{1}{2}\normord{\Big(\partial\psi_t\psi_t + j_t j_t\Big)} \ .
\end{equation}

More generally this correspondence yields a uniform and satisfying interpretation of the chiral algebras we will need in this paper. They are listed in Table~\ref{tab:dictionary}, where we also provide the relevant holonomy groups $\text{G}$, covariantly constant tensors, and our notations for the corresponding worldsheet currents. For instance while Calabi-Yau $n$-folds are widely known to lead to $\mathcal{N}=2$ superconformal symmetry on the worldsheet, their true W-algebra is in fact larger. There is one algebra for each $n\in\mathbb{N}$ and we denote them by $\text{Od}_n$ after Odake \cite{Odake:1988bh}. The $\mathcal{N}=2$ generators are related to the K\"ahler form associated to $\text{G}=\text{U}(n)$; the extra generators are due to the holomorphic $n$-form which is specific to $\text{G}=\text{SU}(n)$. 

\bigskip

\begin{table}[h]
{\footnotesize
\renewcommand\arraystretch{1.3}
\centering
\begin{tabular}{|c|c|c|l|l|c|}
\hline
Dim.\ &
G &
Target space &
Cov.\ const.\ tensors &
Generators of $\Rbag$ {(with weights)} &
Algebra $\Rbag$\\
 &
 &
 &
 &
\& SUSY partners {(with weights)} &
\& c.\ charge
\\\hline\hline
$1$ &
$\mathds{1}$ &
$\mathbb{R}$ or S$^1$ &
$\dd t$ &
$\psi_t$~($\tfrac{1}{2}$) &
$\text{Fr}^1$\\
 &
 &
 &
 &
$j_t=i\partial t$~($1$) &
$c=3/2$
\\\hline
$d$ &
O($d$) &
Riemannian &
$g$ (metric) &
$G$~($\tfrac{3}{2}$) &
$\mathcal{N}=1$\\
 &
 &
 &
 &
$T$~($2$) &
$c=3d/2$
\\\hline
$2n$ &
U($n$) &
K\"ahler &
$g$, $\omega$ (K\"ahler form) &
$G_n$~($\tfrac{3}{2}$), $J^3_n$~($1$) &
$\mathcal{N}=2$\\
 &
 &
 &
 &
$T_n$~($2$), $G^3_n$~($\tfrac{3}{2}$) &
$c=3n$
\\\hline
$2n$ &
SU($n$) &
Calabi-Yau &
$g$, $\omega$, $\Omega$ &
$G_n$~($\tfrac{3}{2}$), $J^3_n$~($1$), $A_n+iB_n$~($\tfrac{n}{2}$) &
$\text{Od}_n$\\
 &
 &
 &
(holom.\ $n$-form) &
$T_n$~($2$), $G^3_n$~($\tfrac{3}{2}$), $C_n+iD_n$~($\tfrac{n+1}{2}$) &
$c=3n$
\\\hline
$7$ &
G$_2$ &
G$_2$ &
$g$, $\varphi$, $*\varphi$ &
$G_7$~($\tfrac{3}{2}$), $P$~($\tfrac{3}{2}$), $X_7$~($2$) &
$\SVseven$\\
 &
 &
holonomy &
(3-form and 4-form) &
$T_7$~($2$), $K$~($2$), $M_7$~($\tfrac{5}{2}$) &
$c=21/2$
\\\hline
$8$ &
Spin(7) &
Spin(7) &
$g$, $\Psi$ (4-form) &
$G_8$~($\tfrac{3}{2}$), $X_8$~($2$) &
$\SVeight$\\
 &
 &
holonomy &
 &
$T_8$~($2$), $M_8$~($\tfrac{5}{2}$) &
$c=12$
\\\hline
\end{tabular}
\caption{Notations and correspondences between covariantly constant tensors on the $\sigma$-model target space with holonomy G and (supersymmetric pairs of) generators of the worldsheet chiral symmetry W-algebra $\Rbag$. The conformal weight of the generators is indicated in parentheses.
}
\label{tab:dictionary}
}
\end{table}

We will only need Od$_n$ for $n=4$, $n=3$ and $n=2$; the latter being actually isomorphic to the small $\mathcal{N}=4$ superconformal algebra at $c=6$. Our conventions for almost all algebras are identical to those of \cite{Fiset:2018huv}, to which we refer for the explicit OPE relations.\footnote{We have added in Table~\ref{tab:dictionary} some subscripts to the different generators in order to distinguish them. Also we should point out that $J^3$ is what was called $J$ in \cite{Fiset:2018huv}.} The only exceptions are the $\SVeight$ algebra, whose OPEs are correctly given in \cite{Shatashvili:1994zw} in the basis that we use, and $\text{Od}_4$. Our main reference for $\text{Od}_4$ is \cite{Figueroa-OFarrill:1996tnk}, but having found some sign errors in that paper, we have produced the OPEs we used in Appendix~\ref{app:Od4}.

The algebras Od$_3$, Od$_4$ and $\SVseven$ are only associative modulo certain singular fields \cite{Odake:1988bh, Figueroa-OFarrill:1996tnk}, respectively
\begin{equation}
N^1_n = \partial A_n - \normord{J^3_n B_n} \ , \qquad
N^2_n = \partial B_n + \normord{J^3_n A_n} \ ,
\end{equation}
for $n=3$ and $n=4$ and
\begin{equation}
N_7 = 4\normord{G_7X_7} -2\normord{P_7K_7}-4\partial M_7-\partial^2 G_7
\end{equation}
for $\SVseven$. More details can be found in \cite{Fiset:2018huv}.

\bigskip

We now give the reasoning behind the diamonds of algebra inclusions discussed in this paper. This is a motivation rather than a derivation, and serves to intuitively appreciate the origin of our main algebraic results presented later.

Notice that our considerations are so far independent of global features of $\mathcal{M}$. Let $\mathcal{U}$ be an open subset in the manifold of interest $\mathcal{M}$. Generally we expect
\begin{equation} \label{eq:SMinSU}
\Rbag(\mathcal{M}) ~\subset~ \Rbag(\mathcal{U}) \ .
\end{equation}
Indeed $\mathcal{U}$ will generically be simpler, more symmetric, than $\mathcal{M}$ so $\Rbag(\mathcal{U})$ should be larger---for example a $\text{U}(n)$-structure may be definable locally but not globally. Moreover it should be possible to realize $\Rbag(\mathcal{M})$, reflecting the global structure, in terms of degrees of freedom of the theory into $\mathcal{U}$. There may however be some flexibility in the way the global structure is reflected in the local theory, leading to some freedom in the embedding \eqref{eq:SMinSU}. For example if $\mathcal{M}$ has $\text{G}=\text{O}(2n)$ and $\mathcal{U}$ has $\text{G}=\text{U}(n)$, then there is a full $\text{S}^1$ orbit's worth of $\Rbag(\mathcal{M})=(\mathcal{N}=1) ~\subset~ (\mathcal{N}=2)= \ \Rbag(\mathcal{U})$, which is actually reflecting the R-symmetry.\footnote{In this example, we are implicitly assuming that $\mathcal{M}$ and $\mathcal{U}$ are of the kind necessary for a quantum CFT to exist in the first place (e.g.\ Ricci-flatness to leading order, etc.).}

\begin{figure}[h]
\begin{center}
\begin{tikzpicture}
\node at (9,0.6) {\rotatebox{45}{$\subset$}};
\node at (11,0.6) {\rotatebox{135}{$\subset$}};
\node at (9,-0.7) {\rotatebox{135}{$\subset$}};
\node at (11,-0.7) {\rotatebox{45}{$\subset$}};
\node at (10,1.25) {$\Rbag(\mathcal{U}\cap\mathcal{V})$};
\node at (8,-0.05) {$\Rbag(\mathcal{U})$};
\node at (12,-0.05) {$\Rbag(\mathcal{V})$};
\node at (10,-1.4) {$\Rbag(\mathcal{M})$};
\end{tikzpicture}
\caption{Diamond of algebra inclusions.}
\label{fig:Diamond}
\end{center}
\end{figure}
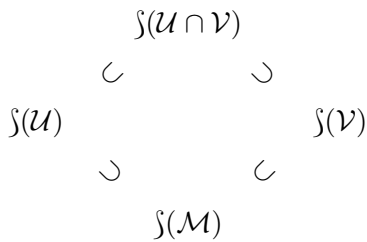

Now let $\mathcal{V}\subset\mathcal{M}$ be another open subset overlapping with $\mathcal{U}$. The same logic yields
\begin{align}
\Rbag(\mathcal{U}) ~\subset~ \Rbag(\mathcal{U}\cap\mathcal{V}) \qquad \text{and} \qquad
\Rbag(\mathcal{V}) ~\subset~ \Rbag(\mathcal{U}\cap\mathcal{V}) \ .
\end{align}
As one transitions from $\mathcal{U}$ to $\mathcal{V}$, it can be expected that the freedom in the embedding \eqref{eq:SMinSU} needs to be restricted. Indeed symmetries emerging locally start being lost if one wants to cover a larger part of $\mathcal{M}$. With the restriction, one achieves that $\Rbag(\mathcal{M})$ also sits in $\Rbag(\mathcal{V})$, and thus we have produced the diamond picture, Figure~\ref{fig:Diamond}. There may remain freedom in the realization of $\Rbag(\mathcal{M})$, but as one covers more and more patches of $\mathcal{M}$, one should be left with a single embedding, since $\Rbag(\mathcal{M})$ are by assumption the symmetries of the theory with configuration space \emph{all} of $\mathcal{M}$.

The ``Connected Sum'' constructions all have in common to be described by only two open patches, so a single diamond reflects the whole geometry $\mathcal{M}=\mathcal{M}_+\cup \mathcal{M}_-$. As a consequence, $\Rbag(\mathcal{M})$ is given precisely by $\Rbag(\mathcal{M}_+) \cap \Rbag(\mathcal{M}_-)$, where the intersection is in $\Rbag(\mathcal{M}_+ \cap \mathcal{M}_-)$.\footnote{$\mathcal{M}_+ \cap \mathcal{M}_-$ is precisely what we have been calling the ``neck'' of the connected sum.} We find moreover in the examples below that geometric transition functions from $\mathcal{M}_+$ to $\mathcal{M}_-$ translate to some algebra automorphism of $\Rbag(\mathcal{M}_+\cap\mathcal{M}_-)$ preserving the $\Rbag(\mathcal{M})$ subalgebra.

The results of \cite{Fiset:2018huv} imply indeed that the $\SVseven$ algebra expected for G$_2$-manifolds is present in two distinct $(\text{Od}_3\oplus\text{Fr}^1)_\pm$ subalgebras of $(\text{Od}_2\oplus\text{Fr}^3)$, see Figure~\ref{fig:TCS}, and thus in their intersection.
The reverse inclusion, unadressed in \cite{Fiset:2018huv}, is perhaps even more interesting, because it informs on worldsheet symmetries of TCS G$_2$-manifolds, which could conceivably be larger than those of a generic G$_2$-manifold. We investigate this in Section~\ref{sec:Num}.

\section{G$_2$ Extra Twisted Connected Sums} \label{sec:G2}

In this section we extend the results of \cite{Fiset:2018huv} to the Extra Twisted Connected Sums (ETCS) of \cite{Crowley:2015ctv, Nordstrom:2018cli, goette2020nuinvariants}. Let us first briefly summarize the geometry.

\subsection{ETCS geometry} \label{sec:G2geom}

The situation for ETCS is very similar to the TCS case shown in Figure~\ref{fig:TCS}(a), since in both cases we glue two Asymptotically Cylindrical Calabi-Yau 3-folds times a circle along a common asymptotic neck region of the form $\text{CY}_2\times \mathbb{T}^2\times \mathbb{R}$. An \emph{Asymptotically Cylindrical (ACyl) Calabi-Yau $n$-fold} ($n$ complex dimensions) has a compact region whose complement is diffeomorphic to a cylinder, here with cross-section a closed Calabi-Yau $(n-1)$-fold times a circle. We represent this asymptotic behaviour by an arrow
\begin{equation}
\text{ACyl CY}_n \longrightarrow \text{CY}_{n-1}\times\Sc^1 \times \R^+ \ .
\end{equation}
In addition, the metric $g_n$ and the K\"ahler and holomorphic volume forms $\omega_n$ and $\Omega_n$ of $\text{CY}_n$ asymptote to those of the cylinder. If we parametrize $\Sc^1$ by $\theta$ and $\R^+$ by $t$, we can write the asymptotic relations between them as follows:
\begin{align}
g_n\longrightarrow g_{n, \infty} &=
g_{n-1} + \dd \theta^2 + \dd t^2 \ , \\
\label{eq:cyintermsofk31}
\omega_n\longrightarrow\omega_{n, \infty} &=\omega_{n-1}+\dd t\wedge\dd\theta \ ,\\
\label{eq:cyintermsofk32}
\Omega_n\longrightarrow\Omega_{n, \infty} &=(\dd\theta-i\dd t)\wedge\Omega_{n-1} \ ,
\end{align}
where the subscript $\infty$ refers to the forms in the limit $t\rightarrow\infty$.

As a first step to generalize the TCS construction we assume there exist cyclic groups $\Gamma_\pm=\mathbb{Z}/k_\pm\mathbb{Z}$
acting diagonally
on the two sides to be glued,  $\text{ACyl~CY}_{3,\pm}\times \text{S}^1_\pm$, in such a way that the Calabi-Yau structure is preserved and that the action is free on the $\text{S}^1_\pm$. We call the latter $\text{S}^1_\pm$ \emph{external circles} and we parametrize them with $\xi_\pm$. Furthermore, in the neck region where the ACyl~CY$_{3,\pm}$ asymptote to  CY$_{2,\pm}\times\text{S}^1_\pm\times\mathbb{R}_\pm$,
we demand that the groups $\Gamma_\pm$ act trivially on CY$_{2,\pm}\times\mathbb{R}_\pm$ and freely on $\text{S}^1_\pm$. We call the latter $\text{S}^1_\pm$ \emph{internal circles} and parametrize them with $\theta_\pm$. Note that, in the asymptotic region, $\Gamma_\pm$ is only acting non-trivially on the torus formed by the internal and external circles.
The quotient $\mathbb{T}^2_\pm/\Gamma_\pm$ is still a torus, however, the group action modifies the original torus lattice, effectively ``twisting'' its structure and changing the length of its sides.

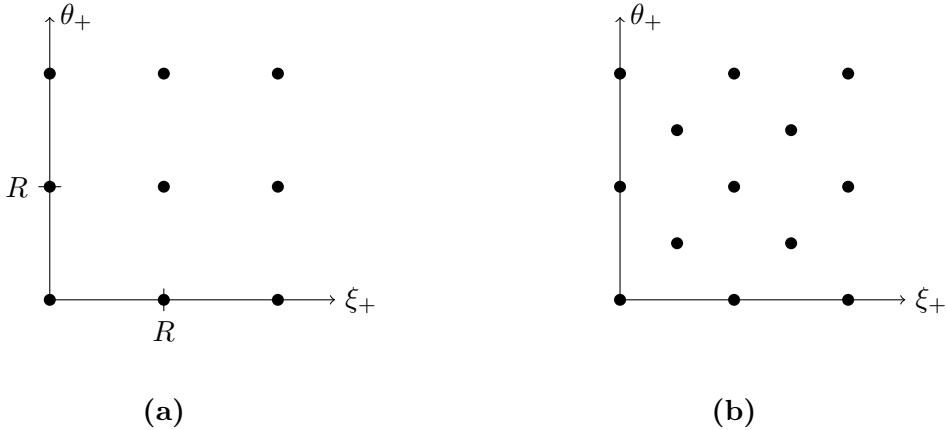
\begin{figure}[h]
\begin{center}
\begin{tikzpicture}[scale=1.5]
\draw[->] (0,0)--(2.5,0) node[right] {$\xi_+$};
\draw[->] (0,0)--(0,2.5) node[right] {$\theta_+$};
\draw[-] (1,0.1)--(1,-0.1) node[below] {$R$};
\draw[-] (0.1,1)--(-0.1,1) node[left] {$R$};
\draw[->] (5,0)--(7.5,0) node[right] {$\xi_+$};
\draw[->] (5,0)--(5,2.5) node[right] {$\theta_+$};
\foreach \y in {0,...,2}{
    \foreach \x in {0,...,2}
    {
    \fill (\x,\y) circle (1.5pt);
    }
    \foreach \x in {5,...,7}
    {
    \fill (\x,\y) circle (1.5pt);
    }
    }
    
\foreach \y in {0.5,1.5}{
    \foreach \x in {5.5,6.5}
    {
    \fill (\x,\y) circle (1.5pt);
    }}

\node at (1,-1) {\textbf{(a)}};
\node at (6,-1) {\textbf{(b)}};
\end{tikzpicture}
\caption{\textbf{(a)} Torus lattice of $\mathbb{T}^2_+$. \textbf{(b)} Torus lattice of $\mathbb{T}^2_+/\Gamma_+$.
}
\label{fig:lattices}
\end{center}
\end{figure}

Let us illustrate these features with an example. The simplest ETCS in \cite{Crowley:2015ctv} involves no quotient on one side of the construction, $\Gamma_-=\lbrace 1 \rbrace$, and a $\mathbb{Z}_2$ quotient on the other side, $\Gamma_+=\lbrace 1,\tau \rbrace$. Note that $\tau$ acts on the CY$_{3,+}$ as an involution which in the asymptotic end performs a rotation of the internal circle by an angle of $\pi$ leaving the rest fixed. The action of $\tau$ on the external circle is also a rotation of angle $\pi$. Take the radius of the internal and external circles
to be of the same length $R$, so that the torus $\mathbb{T}^2_+$ is 
obtained from the square lattice in Figure~\ref{fig:lattices}(a).
The identification under the action of $\tau$
appears in the lattice as a new set of points, see Figure~\ref{fig:lattices}(b). The lattice of
$\mathbb{T}^2_+/\Gamma_+$
still represents a square torus, however the lattice is tilted with respect to the original one and the radius of the circles is now $R/\sqrt{2}$.

The tangent vectors $(\partial_{\theta\pm},\partial_{\xi\pm})$ define an orthonormal frame even after the quotient so we will use them to describe the gluing.
The tori are glued by an orientation-reversing isometry; we call such a map a \emph{torus matching} following \cite{Nordstrom:2018cli}. This was achieved in the TCS case by identifying the internal circle on one side with the external circle on the other side. ETCS require in general a different alignment of the internal and external circles. We assume there exists a torus matching $\mathfrak{t}$ between the tori $\mathbb{T}^2_\pm/\Gamma_\pm$ such that the orthogonal frames are related by
\begin{align}
\label{eq:partialxi} \partial_{\xi-}=\cos\vartheta\partial_{\xi+} +\sin\vartheta\partial_{\theta+} \ ,\\
\label{eq:partialtheta} \partial_{\theta-}=\sin\vartheta\partial_{\xi+} -\cos\vartheta\partial_{\theta+} \ ,
\end{align}
for some $\vartheta\in(0,\pi)$ called the \emph{gluing angle}. This is determined from the tori $\mathbb{T}^2_\pm/\Gamma_\pm$: they are described by the same lattice up to a rotation which essentially
{fixes} the gluing angle. The systematic process to extract this information is described in \cite{goette2020nuinvariants}, but for our purposes it is enough to declare that the lattice of $\mathbb{T}^2_+/\Gamma_+$ is kept fixed whereas the lattice of $\mathbb{T}^2_-/\Gamma_-$ is rotated so that they can be glued together. Then, we can express $(\partial_{\theta-},\partial_{\xi-})$ in terms of $(\partial_{\theta+},\partial_{\xi+})$.
This will later be the key to describe the diamond of algebras for the ETCS case. Note also that the usual TCS corresponds to a gluing angle of $\vartheta=\pi/2$.

\begin{figure}[h]
\begin{center}
\begin{tikzpicture}[scale=1.5]
\foreach \y in {0,...,2}{
    \foreach \x in {0,...,2}
    {
    \fill (\x,\y) circle (1.5pt);
    }    }
    
\foreach \y in {0.5,1.5}{
    \foreach \x in {0.5,1.5}
    {
    \fill (\x,\y) circle (1.5pt);
    }}
    
\draw[->,thick,blue](1,1) -- (1,1.66);
\draw[->,thick,red](1,1) -- (1.66,1);
\draw[->,thick,red](1,1) -- (1.5,1.5);
\draw[->,thick,blue](1,1) -- (1.5,0.5);

\draw[->](1.33,1) arc (0:45:0.33);

\node at (0.75,1.75) {$\partial_{\theta+}$};
\node at (1.5,1.75) {$\partial_{\xi-}$};
\node at (1.75,0.75) {$\partial_{\xi+}$};
\node at (1.5,0.25) {$\partial_{\theta-}$};

\node at (1.5,1.25) {$\vartheta$};
\end{tikzpicture}
\caption{Torus matching between $\mathbb{T}^2_+/\Gamma_+$ and $\mathbb{T}^2_-/\Gamma_-$ with $\vartheta=\pi/4$.}
\label{fig:matching}
\end{center}
\end{figure}
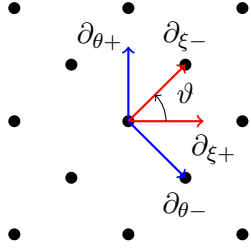

Returning to our example, the torus $\mathbb{T}^2_+/\Gamma_+$ is described by the square lattice in Figure~\ref{fig:lattices}(b) which corresponds to circles of radii $R/\sqrt{2}$. Since the quotient for $\mathbb{T}^2_-/\Gamma_-$ is trivial in our example, we take the radius of the internal and external circles
on this side
to be $R/\sqrt{2}$ (therefore the lattice of $\mathbb{T}^2_-/\Gamma_-$ is Figure~\ref{fig:lattices}(a) with lengths reduced by $\sqrt{2}$). Then the lattices of $\mathbb{T}^2_+/\Gamma_+$ and $\mathbb{T}^2_-/\Gamma_-$ coincide up to a rotation. We can find a torus matching with $\vartheta=\pi/4$ between the lattices, as illustrated in Figure~\ref{fig:matching}.

The K3 surfaces CY$_{2,\pm}$ in the neck region possess hyper-K\"ahler structures given by a triple of closed 2-forms $\omega^I_\pm$, $\omega^J_\pm$ and $\omega^K_\pm$ such that
\begin{equation}
(\omega^I_\pm)^2=(\omega^J_\pm)^2=(\omega^K_\pm)^2\neq 0 \ , \qquad \omega^I_\pm\wedge\omega^J_\pm=\omega^J_\pm\wedge\omega^K_\pm=\omega^K_\pm\wedge\omega^I_\pm=0 \ .
\end{equation}
Calabi-Yau structures determined by Hermitian forms $\omega_\pm$ and holomorphic volume forms $\Omega_\pm$ can then be constructed as
\begin{equation}
\omega_\pm=\omega^I_\pm \ , \qquad \Omega_\pm=\omega^J_\pm+i \omega^K_\pm \ .
\end{equation}
The nontrivial angle in the torus matching implies that a similar 
rotation must be done to the isometry used to glue CY$_{2,+}$ with CY$_{2,-}$ so as to achieve a global G$_2$-structure. We must use a \emph{hyper-K\"ahler matching with angle} $\vartheta$, which we denote by $\mathfrak{r}$ and which satisfies
\begin{equation} \label{eq:hypermatching} \mathfrak{r}^*\omega^K_-=-\omega^K_+ \ ,
\qquad
\mathfrak{r}^*(\omega^I_-+i \omega^J_-)=e^{i\vartheta}(\omega^I_+-i \omega^J_+) \ .
\end{equation}
The TCS hyper-K\"ahler matching is recovered by $\vartheta=\pi/2$, see e.g. \cite[Sect.\ 1.1]{Fiset:2018huv}.

These isometries are used to define a gluing of the two sides of the ETCS construction $\mathcal{M}_\pm=(\text{ACyl~CY}_{3,\pm}\times \text{S}^1_\pm)/\Gamma_\pm$ along the neck region\footnote{Strictly speaking, the ACyl~CY$_{3,\pm}$ must be first truncated at finite distance $t_0$ in the asymptotic direction before being glued. This truncation gives rise to a non-vanishing G$_2$ torsion, going to zero as $t_0\rightarrow\infty$.}
\begin{equation}\label{eq:gluingmapETCS}
F=(-\text{Id}_\mathbb{R})\times\mathfrak{t}\times\mathfrak{r}: \ (\mathbb{R}\times\mathbb{T}^2_+\times\text{CY}_{2,+})/\Gamma_+ \longrightarrow \ (\mathbb{R}\times\mathbb{T}^2_-\times\text{CY}_{2,-})/\Gamma_- \ .
\end{equation}
The resulting manifold after the gluing has a globally-defined G$_2$-structure which is described in the neck region by the same associative 3-form as for the TCS case
\begin{equation} \label{eq:neckg2structure}
\varphi_\pm=\dd\xi_\pm\wedge\omega^I_\pm +\dd\theta_\pm\wedge\omega^J_\pm +\dd t_\pm\wedge\omega^K_\pm+\dd t_\pm\wedge\dd\theta_\pm\wedge\dd\xi_\pm \ .
\end{equation}
The gluing identifies the G$_2$ forms on both sides: $F^*(\varphi_-)=\varphi_+$. The G$_2$-structure has torsion localized around the neck, {however} techniques from analysis show that the G$_2$-structure can be deformed slightly to remove the torsion, obtaining a torsion-free G$_2$-holonomy ETCS manifold.

\subsection{Chiral algebra viewpoint}

Let us now translate this construction to chiral algebras in the worldsheet.
First of all, note that the building blocks of an ETCS possess the same geometric properties as those of a TCS since the quotient does not spoil the Calabi-Yau structure. As a result we
will have
a diamond of inclusions
essentially identical
to the TCS one, see Figure~\ref{fig:TCS}(b), except for how the relevant subalgebras are concretely realized.
Since the neck region
has geometrically the form $\text{CY}_2\times \mathbb{T}^2\times \mathbb{R}$, the
algebra sitting on top of our diamond is $\text{Od}_2\oplus \text{Fr}^3$, consistently with the generalities laid out in Section~\ref{sec:Idea}. That is, as in \cite[Sect.\ 2.2]{Fiset:2018huv}, there is a correspondence between generators of $\text{Od}_2\oplus \text{Fr}^3$ and invariant forms in the neck region, see again Table~\ref{tab:dictionary} for our notations.
To be completely explicit in what follows, we can think of this algebra as being associated to the neck region of $\mathcal{M}_+=(\text{ACyl~CY}_{3,+}\times \text{S}^1_+)/\Gamma_+$, so for example $\dd \theta_+$ is associated with $\psi_\theta$ and $j_\theta$, and so on for the other covariantly constant tensors.

Now the realization of $\text{Od}_3\oplus \text{Fr}^1\subset\text{Od}_2\oplus \text{Fr}^3$ corresponding to the $\mathcal{M}_+$ side is precisely the one found in \cite{Fiset:2018huv}.\footnote{In order to perform all the computations involving operator algebras in this work, we used the package \emph{OPEdefs} by Thielemans \cite{Thielemans:1994er}.} This is because the quotient by $\Gamma_+$ preserves the Calabi-Yau structure and the asymptotic description of the different forms. Explicitly the generators of $(\text{Od}_3\oplus \text{Fr}^1)_+$ are given by $\psi_\xi$ and
\begin{equation} \label{eq:Od3inOd2F2}
\begin{split}
G_3 &= G_2 + G_\theta + G_t \ , \\
J^3_3 &= J^3_2 \ + \normord{\psi_t\psi_\theta} \ , \\
A_3+iB_3 &= \ \normord{(\psi_\theta-i\psi_t)(A_2+iB_2)} \ ,
\end{split}
\end{equation}
along with their supersymmetric partners, see Table~\ref{tab:dictionary}, which can all be reconstructed from \eqref{eq:Od3inOd2F2}.

Moreover the expression \eqref{eq:neckg2structure} of the G$_2$-structure in the neck region  is the same as in the TCS case, so the realisation of SV$^{\text{G}_2}\subset\text{Od}_2\oplus \text{Fr}^3$ \cite{Figueroa-OFarrill:1996tnk} used in \cite{Fiset:2018huv} also applies to this ETCS case.
Explicitly\footnote{Note that it is sufficient to specify the $\text{SV}^{G_2}$ generators $G_7$ and $P$, as the others can be deduced from them from operator product expansions.}
\begin{equation}
G_7 = G_3 + G_\xi \ , \qquad
P = A_3 \ + \normord{J^3_3 \psi_\xi} \ ,
\end{equation}
so combining with \eqref{eq:Od3inOd2F2},
\begin{equation} \label{eq:SV7inOd2F3}
G_7 = G_2 + G_\theta + G_\xi + G_t \ , \qquad
P = \ \normord{\psi_\theta A_2} +\normord{\psi_t B_2} + \normord{J^3_2 \psi_\xi} + \normord{\psi_t\psi_\theta\psi_\xi} \ .
\end{equation}

The other subalgebra $(\text{Od}_3\oplus \text{Fr}^1)_- \subset\text{Od}_2\oplus \text{Fr}^3$ corresponding to $\mathcal{M}_-=(\text{ACyl~CY}_{3,-}\times \text{S}^1_-)/\Gamma_-$ is obtained from \eqref{eq:Od3inOd2F2} by applying the following automorphism of $\text{Od}_2\oplus \text{Fr}^3$, which is inferred from the geometric gluing map, eq.~\eqref{eq:gluingmapETCS}:
\begin{equation} \label{eq:GluingETCS}
\begin{split}
G_2 &\longmapsto G_2 \\
\begin{pmatrix}
J^3_2 \\ A_2
\end{pmatrix} &\longmapsto
\begin{pmatrix}
\cos\vartheta & \sin\vartheta\\
\sin\vartheta & -\cos\vartheta
\end{pmatrix}
\begin{pmatrix}
J^3_2 \\ A_2
\end{pmatrix}
\qquad\qquad
\begin{pmatrix}
\psi_\xi \\ \psi_\theta
\end{pmatrix} \longmapsto
\begin{pmatrix}
\cos\vartheta & \sin\vartheta\\
\sin\vartheta & -\cos\vartheta
\end{pmatrix}
\begin{pmatrix}
\psi_\xi \\ \psi_\theta
\end{pmatrix}
\\
B_2 &\longmapsto -B_2
\qquad\qquad\qquad
\qquad\qquad\qquad
\quad\ \,
\psi_t \longmapsto -\psi_t
\end{split}
\end{equation}
(with the same action on supersymmetric partners). Indeed applying this to the expressions \eqref{eq:Od3inOd2F2} generates $(\text{Od}_3 \oplus \text{Fr}^1)_-$ in terms of  $\cos\vartheta\psi_\xi + \sin\vartheta\psi_\theta$ and
\begin{equation} \label{eq:Od3inOd2F2Prime}
\begin{split}
G_3 &= G_2 + \sin^2\vartheta G_\xi +\cos^2\vartheta G_\theta -\sin\vartheta\cos\vartheta (\normord{j_\xi\psi_\theta}+\normord{j_\theta\psi_\xi}) + G_t \ ,
\\
J^3_3 &= \cos\vartheta (J^3_2 +\normord{\psi_t\psi_\theta})+\sin\vartheta(A_2-\normord{\psi_t\psi_\xi}) \ ,
\\
A_3+iB_3 &= \;
\normord{(\sin\vartheta\psi_\xi-\cos\vartheta\psi_\theta+i\psi_t)(\sin\vartheta J^3_2-\cos\vartheta A_2 - iB_2)} \ .
\end{split}
\end{equation}
The ordinary TCS automorphism of \cite{Fiset:2018huv} is recovered from \eqref{eq:GluingETCS} by setting $\vartheta=\pi/2$.

It can be checked that, for any $\vartheta$, the map \eqref{eq:GluingETCS} leaves SV$^{\text{G}_2}$ in \eqref{eq:SV7inOd2F3} invariant. This shows that the diamond of inclusions, Figure~\ref{fig:TCS}, is indeed correct: SV$^{\text{G}_2}$ sits in the intersection of $(\text{Od}_3 \oplus \text{Fr}^1)_+$ with $(\text{Od}_3 \oplus \text{Fr}^1)_-$ inside $\text{Od}_2 \oplus \text{Fr}^3$ for any $\vartheta$.

\bigskip

We stress that these various statements rely on a careful treatment of null vectors. The OPEs of $\text{Od}_3 \oplus \text{Fr}^1$ are only satisfied by \eqref{eq:Od3inOd2F2} upon quotienting by \cite{Odake:1988bh}
\begin{equation}
N^1_3 = \partial A_3 \; -\normord{J^3_3 B_3} \ , \qquad
N^2_3 = \partial B_3 \; +\normord{J^3_3 A_3} \ ,
\end{equation}
where $J^3_3$, $A_3$, $B_3$ are given by \eqref{eq:Od3inOd2F2}. Similarly the image $(\text{Od}_3 \oplus \text{Fr}^1)_-$ \eqref{eq:Od3inOd2F2Prime} under the gluing automorphism \eqref{eq:GluingETCS} is only valid up to the image of $N^1_3$, $N^2_3$. When regarded as elements of $\text{Od}_2 \oplus \text{Fr}^3$, the fields $N^1_3$, $N^2_3$, and their images under \eqref{eq:GluingETCS}, all descend from the null vectors
\begin{equation}
N^1_2 = \partial A_2 \; -\normord{J^3_2 B_2} \ , \qquad
N^2_2 = \partial B_2 \; +\normord{J^3_2 A_2} \ .
\end{equation}
We always assume that null vectors of $\text{Od}_2\oplus \text{Fr}^3$ are quotiented out. This also ensures that the null vector $N$ modulo which the $\SVseven$ subalgebra is associative, see \cite{Figueroa-OFarrill:1996tnk}, is indeed zero.

It is worth stressing that the diamond diagram is well-defined for any value of the gluing angle $\vartheta$. ETCS have been constructed so far only for a discrete set of gluing angles. Our results show that, at least from the worldsheet algebra perspective, there is no reason to exclude ETCS with somewhat more general gluing angles.

\section{Spin(7) Generalized Connected Sums} \label{sec:Spin7}

\subsection{GCS geometry} \label{sec:Spin7Geom}

\begin{figure}[h] 
\begin{center}
\begin{tikzpicture}
\draw [black]
(2,0.5) to (1,0.5)
to[out=180, in=-10] (-0.5,0.75)
to[out=170, in=90] (-1.5,0)
to[out=-90, in=190] (-0.5,-0.75)
to[out=10, in=180] (1,-0.5) to (2,-0.5);
\draw [black]
(0.75,0.4) to (1.75,0.4)
to[out=0, in=190] (3.25,0.65)
to[out=10, in=90] (4.25,-0.1)
to[out=-90, in=-10] (3.25,-0.85)
to[out=170, in=0] (1.75,-0.6) to (0.75,-0.6);
\node at (1.375,0.8) {$\overbrace{\qquad\qquad\qquad}$};
\node at (1.375,-1.0) {$\underbrace{\qquad\qquad\qquad\qquad\qquad\qquad\qquad}$};
\node at (1.375,-2.5) {\textbf{(a)}};
\node at (10,-2.5) {\textbf{(b)}};
\node at (9,0.6) {\rotatebox{45}{$\subset$}};
\node at (11,0.6) {\rotatebox{135}{$\subset$}};
\node at (9,-0.7) {\rotatebox{135}{$\subset$}};
\node at (11,-0.7) {\rotatebox{45}{$\subset$}};

\node at (1.375,1.25) {$\text{CY}_3\times \text{S}^1\times \mathbb{R}$};
\node at (-0.5,0) {$\text{CY}_4$};
\node at (3.2,-0.1) {$M_7 \times \text{S}^1$};
\node at (1.375,-1.5) {Spin(7) GCS};

\node at (10,1.25) {$\text{Od}_3\oplus \text{Fr}^2$};
\node at (8.3,-0.05) {$\text{Od}_4$};
\node at (12,-0.05) {$\SVseven \oplus \text{Fr}^1$};
\node at (10,-1.4) {$\SVeight$};
\end{tikzpicture}
\caption{\textbf{(a)} Sketch of a compact 8-dimensional Spin(7)-holonomy manifold obtained as Generalized Connected Sum (GCS).{ $M_7$ is an ACyl G$_2$-manifold.} \textbf{(b)} Diamond of algebra inclusions corresponding to a $\sigma$-model whose target space is a GCS.}
\label{fig:GCS}
\end{center}
\end{figure}
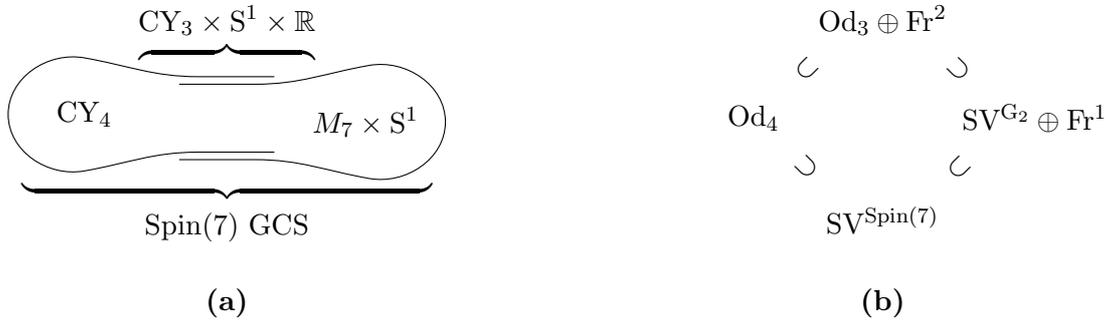

Strong evidence for the existence of a generalization of G${}_2$ TCS for the case of Spin(7)-manifolds, called ``Generalized Connected Sum'' (GCS), was provided in \cite{Braun:2018joh}. From specific examples (Joyce orbifolds of $\mathbb{T}^8$), the authors show that in this case the two manifolds to be glued along the neck region must be different, see Figure~\ref{fig:GCS}(a).

\noindent One of the manifolds is an ACyl Calabi-Yau 4-fold $\mathcal{M}_+$, see Section~\ref{sec:G2geom}. We sometimes use $+$ to distinguish any object relative to it.

For the other half of the construction we need to introduce a similar notion. An \emph{ACyl G$_2$-manifold $M_7$}, see e.g.\ \cite{MR2442130, MR2721660}, is a non-compact manifold of holonomy G${}_2$ with a compact region whose complement is diffeomorphic to a closed Calabi-Yau 3-fold times the positive real line. We represent this by
\begin{equation}
\text{ACyl }M_7 \longrightarrow \text{CY}_3 \times \R^+ \ .
\end{equation}
In addition, the associative 3-form $\varphi$ and coassociative 4-form $*\varphi$ of $M_7$ are asymptotic to the ones that can be constructed from the Calabi-Yau times the real line; parametrizing $\mathbb{R}$ by $t$, we have
\begin{align}
\label{eq:g2intermsofcy1}
\varphi\longrightarrow \varphi_\infty &=\Re\Omega_3 -\dd t\wedge\omega_3 \ ,\\
\label{eq:g2intermsofcy2}
*\varphi\longrightarrow*\varphi_\infty &=\frac{1}{2}\omega_3\wedge\omega_3+\dd t\wedge\Im\Omega_3 \ ,
\end{align}
where again the subscript $\infty$ refers to the forms in the asymptotic limit and where $\Re$ and $\Im$ are the real and imaginary parts.

The manifolds we are going to glue together are, on the one hand, an ACyl CY$_4$ $\mathcal{M}_+$ and, on the other hand, the product of a circle with an ACyl G$_2$ manifold $\mathcal{M}_- = M_7 \times \text{S}^1$. The manifolds $\mathcal{M}_+$ and $\mathcal{M}_-$ can be glued along their asymptotic ends, which are compatible. Introducing a boundary at $t=t_0+1$ (on both sides), we can define the gluing map along the interval $I=[t_0,t_0+1]$:
\begin{align}
\label{eq:gluingmapSpin7}
\begin{split}
F_{t_0}: 
\text{CY}_{3, +} \times \text{S}^1_+ \times I_+
&\longrightarrow 
\text{CY}_{3, -} \times \text{S}^1_- \times I_-
\\
\big(z,~\theta,~t\big) &\longmapsto \big(\phi(z),~-\theta,~2t_0+1-t\big) \ ,
\end{split}
\end{align}
where $\phi$ is a biholomorphic map between the Calabi-Yau 3-folds such that the holomorphic volume form changes global sign, that is, $F_{t_0}^*(\Omega_{3,-})=-\Omega_{3,+}$. {With our conventions we also need to reverse the sign in the identification of the circles in order to have the correct gluing.}

Next we have to specify a Spin(7)-structure on the whole manifold, which is done with a Cayley 4-form $\Psi$ \cite{Joyce2007}. First of all, since $\mathcal{M}_+$ has holonomy SU(4) and $\text{SU(4)}\subset\text{Spin(7)}$, we can define a torsion-free Spin(7)-structure on $\mathcal{M}_+$ via
\begin{equation}
\label{eq:psiintermsofcy4}
\Psi_+=\Re\Omega_{4,+}+\frac{1}{2}\omega_{4,+}\wedge\omega_{4,+} \ .
\end{equation}
In the asymptotic neck region $\omega_{4,+}$ and $\Omega_{4,+}$ decompose according to \eqref{eq:cyintermsofk31} and \eqref{eq:cyintermsofk32}. Thus, the Spin(7)-structure takes the form
\begin{equation}
\label{psi+}
\Psi_{\infty,+}=\dd\theta_+\wedge\Re\Omega_{3,+}+\dd t_+\wedge\Im\Omega_{3,+}+\frac{1}{2}\omega_{3,+}\wedge\omega_{3,+}+\dd t_+\wedge\dd\theta_+\wedge\omega_{3,+} \ .
\end{equation}
On the other hand, the manifold $\mathcal{M}_-$ has holonomy G${}_2$ and ${\rm G}_2\subset\text{Spin(7)}$, so that we can construct a torsion-free Spin(7)-structure as follows
\begin{equation}
\label{eq:psiintermsofg2circle}
\Psi_-=\dd\theta_-\wedge\varphi_-+*\varphi_- \ .
\end{equation}
In the asymptotic region, the associative and coassociative forms can be decomposed as in \eqref{eq:g2intermsofcy1} and \eqref{eq:g2intermsofcy2}, so the Spin(7)-structure takes the form
\begin{equation}
\label{psi-}
\Psi_{\infty,-}=\dd\theta_-\wedge\Re\Omega_{3,-}+\dd t_-\wedge\Im\Omega_{3,-}+\frac{1}{2}\omega_{3,-}\wedge\omega_{3,-}+\dd t_-\wedge\dd\theta_-\wedge\omega_{3,-} \ .
\end{equation}
The diffeomorphism $F_{t_0}$ identifies the Spin(7)-structures $\Psi_-$ and $\Psi_+$ along the gluing, so that the resulting manifold has a global Spin(7)-structure. As in the G${}_2$ TCS case, the structure is torsion-free except around the gluing region (because of the truncation at finite $t_0$). It is believed that, analogously to the G$_2$ case, for large enough $t_0$, a small deformation of the structure can be found such that the resulting manifold has torsion-free Spin(7)-structure, i.e. Spin(7)-holonomy.

\subsection{Chiral algebra viewpoint} \label{sec:Spin7Alg}

We now describe how the GCS construction is reflected in the worldsheet chiral algebras, see Figure~\ref{fig:GCS}(b). The neck region where the open patches of a GCS overlap is given by $\text{CY}_3\times\text{S}^1\times\mathbb{R}$ so the top algebra is given by $\text{Od}_3\oplus \text{Fr}^2$. Now, one of the open patches has a $\text{CY}_4$ structure so we expect an $\text{Od}_4$ subalgebra in one of the sides of the diamond, whereas the other open patch has the structure of a G$_2$-manifold times a circle so we expect a $\SVseven\oplus\text{Fr}^1$ on that side. Since the manifold has holonomy Spin(7) after the gluing, we expect a $\SVeight$ algebra at the bottom of the diagram.

Let us now provide these inclusions explicitly. First of all, we describe $\SVseven\oplus\text{Fr}^1\subset\text{Od}_3\oplus \text{Fr}^2$. The $\text{Fr}^1$ part corresponding geometrically to the circle is trivially identified, so we only require a realization of $\SVseven\subset\text{Od}_3\oplus \text{Fr}^1$. This was already found in \cite{Figueroa-OFarrill:1996tnk} and in fact constitutes the bottom inclusion of the TCS diagram in Figure~\ref{fig:TCS}(b). To be completely explicit, the ACyl G$_2$-manifold definition \eqref{eq:g2intermsofcy1}--\eqref{eq:g2intermsofcy2} suggests the following ansatz, where the subindex $t$ stands for the $\text{Fr}^1$ corresponding to the real line $\mathbb{R}$:
\begin{equation}\label{eq:g2frsubod3fr2}
\begin{split}
&G_7=G_3 + G_t \ , \qquad
P= A_3 \, -\normord{\psi_t J^3_3} \ , \qquad
X_7=\frac{1}{2}\normord{J^3_3J^3_3}+\normord{\psi_t B_3}- \frac{1}{2}\normord{\partial\psi_t\psi_t} \ ,
\end{split}
\end{equation}
These operators and their descendants indeed satisfy the $\SVseven$ OPE relations up to null vectors of Od$_3$.

A realization of $\SVeight\subset\SVseven\oplus\text{Fr}^1$ was already found in \cite{Shatashvili:1994zw, Gepner:2001px}. Explicitly, an ansatz is given by \eqref{eq:psiintermsofg2circle} and denoting the fields from the $\text{Fr}^1$ associated to the circle with a subindex $\theta$ we find
\begin{equation}\label{eq:spin7subg2fr}
\begin{split}
&G_8=G_7+G_\theta \ , \qquad 
X_8=-\left(\normord{\psi_\theta P}+X_7\right)+\frac{1}{2}\normord{\partial\psi_\theta\psi_\theta}.
\end{split}
\end{equation}
These operators and their descendants satisfy the $\SVeight$ algebra OPEs up to null vectors of $\SVseven$.

We now turn our attention to the other side of the diamond: $\text{Od}_4\subset\text{Od}_3\oplus \text{Fr}^2$. In order to find a realization in $\text{Od}_3\oplus \text{Fr}^2$, geometry again provides inspiration. From the asymptotic formulae \eqref{eq:cyintermsofk31} and \eqref{eq:cyintermsofk32}, we write
\begin{equation}\label{eq:od4subod3fr2}
\begin{split}
G_4 &=G_3+G_\theta+G_t \ ,\\
J^3_4 &=J^3_3\,+\normord{\psi_t\psi_\theta} \ ,\\
A_4+iB_4 &=\,\normord{(\psi_\theta-i\psi_t)(A_3+iB_3)} \ .
\end{split}
\end{equation}
We find that these operators and their descendants indeed satisfy the Od${}_4$ OPEs up to null fields of $\text{Od}_3$.\footnote{Moreover the null fields $N^1_4$, $N^2_4$ modulo which Od${}_4$ is associative, see Section~\ref{sec:Idea}, are indeed null since they can be rewritten in terms of null fields of Od${}_3$:
$
N^1_4=\,\normord{\psi_\theta N^1_3}+\normord{\psi_t N^2_3} $ and
$
N^2_4=\,\normord{\psi_\theta N^2_3}-\normord{\psi_t N^1_3}
$.
}

The last relation of the diagram that we must give is $\SVeight\subset\text{Od}_4$. The existence of this subalgebra was already shown in \cite{Figueroa-OFarrill:1996tnk}; it again follows from geometric considerations, in particular \eqref{eq:psiintermsofcy4}:\footnote{Note the overall minus sign in $X_8$ as compared to the geometric formula \eqref{eq:psiintermsofcy4} for {$\Psi$}. This is because of an unfortunate convention in the field theory side. See \cite[Section 2]{Fiset:2019ecu} for other instances of the same mismatch.}
\begin{equation}\label{eq:spin7subod4}
\begin{split}
&G_8=G_4 \ ,\qquad 
X_8=-\Big( A_4+\frac{1}{2}\normord{J^3_4J^3_4}\Big) \ .
\end{split}
\end{equation}
Again these operators and their descendants satisfy the $\SVeight$ OPE relations up to the Od${}_4$ null fields.

There is one last check we have to perform. We have described two embeddings of the $\SVeight$ algebra inside $\text{Od}_3\oplus \text{Fr}^2$, one of them via an $\text{Od}_4$ subalgebra and the other via a $\SVseven\oplus\text{Fr}^1$ subalgebra. For the diamond to hold we must ensure that these embeddings are precisely the same. This is indeed the case: once the $\text{Od}_4$ and $\SVseven\oplus\text{Fr}^1$ generators are rewritten in terms of $\text{Od}_3\oplus \text{Fr}^2$ operators we find that the generators of both $\SVeight$ algebras are exactly the same.

\bigskip

Recall that in the (E)TCS case the gluing morphism \eqref{eq:GluingETCS} inferred from geometry mapped the two sides of the diamond to one another. In the GCS case however the sides of the diamond have non-isomorphic subalgebras so we find a different behaviour. The gluing map \eqref{eq:gluingmapSpin7} suggests the following automorphism of $\text{Od}_3\oplus \text{Fr}^2$:
\begin{equation}
\label{eq:spin7gluingmorphism}
\begin{split}
G_{3} &\longmapsto G_{3} \ ,
\qquad\qquad\qquad
\psi_\theta \longmapsto-\psi_\theta \ ,\\
J^3_{3} &\longmapsto J^3_{3} \ ,
\qquad\qquad\qquad~
\psi_t \longmapsto-\psi_t \ ,\\
A_{3}+iB_3 &\longmapsto -(A_{3}+iB_3) \ .\\
\end{split}
\end{equation}
This map induces automorphisms of the algebras at the sides of the diamond: the identity map on the Od$_4$ side, and the following automorphism on the $\SVseven\oplus\text{Fr}^1$ side:
\begin{equation}
G_7 \longmapsto G_7 \ ,
\qquad
P \longmapsto -P \ ,
\qquad
\psi_\theta \longmapsto - \psi_\theta \ .
\end{equation}
Just like in the (E)TCS case, the Shatashvili-Vafa algebra at the bottom of the diamond is left invariant by the gluing morphism \eqref{eq:spin7gluingmorphism}.

\bigskip

This realization of the diamond of subalgebras heavily relies on the geometry of the GCS construction. A natural question is what happens to the diamond {in the case when different conventions can be} chosen for the geometric structures. Different conventions change the explicit embeddings of the subalgebras inside $\text{Od}_3\oplus \text{Fr}^2$. For example, rotating by a phase $A_3+iB_3$ (akin to the holomorphic volume form) is an automorphism of $\text{Od}_3$ producing from \eqref{eq:g2frsubod3fr2} a U(1)-family of $\SVseven\oplus\text{Fr}^1$ subalgebras.
The $\SVeight$ embedding is also modified. Nevertheless it is reassuring that we always find a diamond structure with a single $\SVeight$ in the intersection of the algebras in the lateral tips. Once conventions are fixed we find a single diamond of subalgebras, as we expected.

\subsection{Automorphisms and mirror symmetry} \label{sec:Spin7Aut}

Mirror symmetry for exceptional holonomy manifolds was first suggested in \cite{Shatashvili:1994zw}, see also \cite{Papadopoulos:1995da}, and later examined through Joyce orbifold examples \cite{Acharya:1996fx, Acharya:1997rh, Gaberdiel:2004vx, Chuang:2004th} and examples of the form $\big(\text{S}^1\times \text{CY}_3\big)/\mathbb{Z}_2$ \cite{Partouche:2000uq, Salur:2007ev}.
 The case of TCS manifolds was fruitfully explored in \cite{Braun:2017ryx} and \cite{Braun:2017csz}, whereas GCS mirror symmetry was recently adressed in \cite{Braun:2019lnn}. Some of the TCS mirror maps were interpreted in terms of chiral algebras in \cite{Fiset:2018huv} and in this section we perform a similar study for GCS manifolds, using the diamond of chiral algebras. {We also exploit that some Joyce orbifolds admit a GCS description in order to compare and propose new mirror constructions.}

A mirror symmetry map may alter drastically the geometry of the target manifold, but the $\sigma$-model theory is preserved; {in particular} the mirror map corresponds to an automorphism of the chiral algebra. Furthermore, for the case of GCS manifolds it is natural to look for mirrors which also possess a GCS structure, as was the case in \cite{Braun:2019lnn}. We have shown that this geometric structure is encoded in the diamond of chiral algebras, see Figure~\ref{fig:GCS}(b), so a mirror map respecting the GCS structure has to correspond to an automorphism of the top algebra $\text{Od}_3\oplus \text{Fr}^2$ preserving the diamond. In particular, it has to reduce to automorphisms of the algebras at the lateral tips and at the bottom of the diamond.

We {performed} a systematic search for these automorphisms. An important observation is that only two automorphisms of $\SVeight$ exist:\footnote{We demand that the Virasoro subalgebra generated by $T_8$ and $G_8$ should also be preserved.} the identity map and the parity map {$(-1)^F$ (where $F$ acts as $0$ on bosons and $1$ on fermions)}. It turns out we do not miss interesting information by restricting ourselves to automorphisms that reduce to the identity on $\SVeight$.\footnote{Any automorphism reducing to the identity on $\SVeight$ can be composed with $(-1)^F$ on $\text{Od}_3\oplus \text{Fr}^2$ to produce an automorphism reducing to $(-1)^F$ on $\SVeight$, and vice versa.} Moreover it is not too hard to see that the automorphisms must act diagonally on $\text{Od}_3\oplus \text{Fr}^1_t \oplus \text{Fr}^1_\theta$.\footnote{This follows from the condition that the Virasoro algebra of $\text{Od}_3\oplus \text{Fr}^2$ has to be preserved by the automorphism, and that the map should reduce to an automorphism of the lateral algebra $\SVseven\oplus\text{Fr}^1$.}

There are only four automorphisms of $\text{Od}_3\oplus\text{Fr}^2$ which satisfy these constraints. One of them is the identity, which we denote by $\textbf{A}_0$. The nontrivial automorphisms $\textbf{A}_1$, $\textbf{A}_2$, $\textbf{A}_3$ and their restriction to the subalgebras of the diamond are described in Table~\ref{tab:automorphisms} below. {We remark that the $\mathbf{A}_i$ form the group $\mathbb{Z}_2^2$ under composition.} In the table, $\textbf{Ph}^\pi$ is defined as
\begin{equation}
\textbf{Ph}^\pi ~:~
A_n+iB_n \longmapsto -(A_n+iB_n) \ , \qquad
C_n+iD_n \longmapsto -(C_n+iD_n) \ ,
\end{equation}
(with the other generators invariant) and it is interpreted geometrically as a phase rotation by $\pi$ of the Calabi-Yau volume form. The other boldface maps will be explained shortly.

\renewcommand{\arraystretch}{1.2}
\begin{table}[h]
\begin{center}
\begin{tabular}{ |c|c|c|c|c| } 
\hline
 $\text{Automorphisms}$  & $\textbf{A}_0$ & $\textbf{A}_1$ & $\textbf{A}_2$ & $\textbf{A}_3$ \\
 \hline
 \hline
 $\text{Od}_3\oplus \text{Fr}^2$ & $\textbf{Id}$ & $\textbf{Ph}^\pi\circ\textbf{T}_t\circ\textbf{T}_\theta$ & $\textbf{M}\circ\textbf{T}_t$ & $\textbf{M}\circ\textbf{Ph}^\pi\circ\textbf{T}_\theta$ \\
 \hline
$\SVseven\oplus\text{Fr}^1$ & $\textbf{Id}$ & $\textbf{GK}\circ\textbf{T}_\theta$ & $\textbf{Id}$ & $\textbf{GK}\circ\textbf{T}_\theta$ \\ 
\hline
$\text{Od}_4$ & $\textbf{Id}$ & $\textbf{Id}$ & $\textbf{M}$ & $\textbf{M}$ \\ 
\hline
$\SVeight$ & $\textbf{Id}$ & $\textbf{Id}$ & $\textbf{Id}$ & $\textbf{Id}$ \\
\hline
\end{tabular}
\end{center}
\caption{Candidates to mirror automorphisms and their action on the algebras of the diamond.}
\label{tab:automorphisms}
\end{table}

Recall that T-duality along a direction $\text{S}^1_\theta$ is accompanied on the worldsheet by the automorphism
\begin{equation}
\label{eq:tdualityautomorphism}
\textbf{T}_\theta ~:~
\psi_\theta \longmapsto -\psi_\theta \ , \qquad
j_\theta \longmapsto -j_\theta \ ,
\end{equation}
acting on, say, the left-moving $\text{Fr}^1_\theta$ (but not on the right-moving $\overline{\text{Fr}^1_\theta}$).
A direction $\R_t$ gives rise to a worldsheet algebra $\text{Fr}^1_t$, so by a map $\textbf{T}_t$ we mean the analogous of the worldsheet automorphism \eqref{eq:tdualityautomorphism} acting on the currents $(\psi_t,j_t)$\footnote{When the target manifold is a Joyce orbifold the map $\textbf{T}_t$ arises from a T-duality along the $t$ direction on the underlying torus. In the general case, the global geometric interpretation of $\textbf{T}_t$ is not clear even though the automorphism is perfectly well-defined. This was to be expected because the chiral algebra only captures the local behaviour of the target manifold: for example a line and a circle both give rise to the same worldsheet currents and the chiral algebras can not be told apart.}. Mirror symmetry for Joyce orbifolds is essentially a combination of $\textbf{T}$ maps, as we now recall.

For the G$_2$ Joyce orbifolds two different maps were described in \cite{Acharya:1997rh}: one is obtained by T-dualizing along {associative} $\mathbb{T}^3$ fibres and the other one by T-dualizing along {coassociative} $\mathbb{T}^4$ fibres. We denote them by $\mathcal{T}^3$ and $\mathcal{T}^4$ respectively. In \cite{Gaberdiel:2004vx} it was shown that $\mathcal{T}^4$ leads to the identity automorphism of $\SVseven$, and that $\mathcal{T}^3$ leads to the automorphism
\begin{equation}
\textbf{GK} ~:~
P \longmapsto -P \ , \qquad
K \longmapsto -K \ ,
\end{equation}
of $\SVseven$, which we call the Gaberdiel-Kaste mirror map. These considerations are G$_2$ analogues of the familiar SYZ conjecture, wherein T-duality along {a supersymmetric $\mathbb{T}^n$ fibration of a} $\text{CY}_n$ gives rise to mirror symmetry and the following worldsheet automorphism of Od$_n$:
\begin{equation}
\textbf{M} ~:~ 
\big(J^3_n,G^3_n,B_n,D_n\big)
\longmapsto
\big(-J^3_n,-G^3_n,-B_n,-D_n\big)
\ .
\end{equation}

For Joyce orbifolds of Spin(7) holonomy, {we give a detailed account in the next section.}

\subsubsection{Spin(7) Joyce orbifolds}

\cite{Acharya:1997rh} describes a single {type of} mirror map in that case, obtained by T-dualizing along {supersymmetric} $\mathbb{T}^4$ fibres. For {any} given Joyce orbifold there are 14 such fibrations, corresponding to toroidal fibres which are calibrated by the Cayley 4-form 
\begin{align} \label{eq:spin7fourform}
    \Psi=\; &\dd x^{1234} +\dd x^{1256} +\dd x^{1278} +\dd x^{1357} -\dd x^{1368} -\dd x^{1458} -\dd x^{1467} \nonumber \\
    & -\dd x^{2358} -\dd x^{2367} -\dd x^{2457} +\dd x^{2468} +\dd x^{3456} +\dd x^{3478} +\dd x^{5678} \ ,
\end{align}
where $\dd x^{ijkl}$ stands for $\dd x^i \wedge\dd x^j \wedge\dd x^k \wedge\dd x^l$.
The combinations of {four} T-dualities that give a mirror map can be read off the terms of the 4-form:
\begin{align} \label{eq:listofTdualities}
    \lbrace &(1, 2, 3, 4), (1, 2, 5, 6), (1, 2, 7, 8), (1, 3, 5, 7), (1, 3, 6, 8), (1, 4, 5, 8), (1, 4, 6, 7), \nonumber \\
    &(2, 3, 5, 8), (2, 3, 6, 7), (2, 4, 5, 7), (2, 4, 6, 8), (3, 4, 5, 6), (3, 4, 7, 8), (5, 6, 7, 8) \rbrace .
\end{align}

As shown in \cite{Braun:2018joh}, there are some Joyce orbifolds which also admit a GCS description, so a natural step for us is to consider what automorphisms of the diamond are generated by the mirror maps \eqref{eq:listofTdualities} in these particular orbifolds (see \cite{Chuang:2004th, Braun:2019lnn} for partial results). We thus consider Spin(7) orbifolds of the form $\mathbb{T}^8/\mathbb{Z}_2^4$ \cite{MR1383960}. The coordinates $x^i$ on the torus range from 0 to 1 and the action of the generators {$\alpha, \beta, \gamma, \delta$} of the quotient group are described in Table~\ref{tab:discreteaction}. We focus on four particular orbifolds, described in Table~\ref{tab:exampleslist}.

{\renewcommand{\arraystretch}{1.2}
\begin{table}[h]
\begin{center}
\begin{tabular}{ |c||c|c|c|c|c|c|c|c| } 
\hline
   & $x^1$ & $x^2$ & $x^3$ & $x^4$ & $x^5$ & $x^6$ & $x^7$ & $x^8$ \\
 \hline
 \hline
 $\alpha$ & $-$ & $-$ & $-$ & $-$ & $+$ & $+$ & $+$ & $+$ \\
 \hline
$\beta$ & $+$ & $+$ & $+$ & $+$ & $-$ & $-$ & $-$ & $-$ \\ 
\hline
$\gamma$ & $c_1-$ & $c_2-$ & $+$ & $+$ & $c_5-$ & $c_6-$ & $+$ & $+$ \\ 
\hline
$\delta$ & $d_1-$ & $+$ & $d_3-$ & $+$ & $d_5-$ & $+$ & $d_7-$ & $+$ \\
\hline
\end{tabular}
\end{center}
\caption{Action of $\mathbb{Z}^4_2$ on $\mathbb{T}^8$. We need to specify the parameters $c_j$ and $d_k$, which are allowed to take the values 0 or $\frac{1}{2}$. This is done in Table~\ref{tab:exampleslist} for four different orbifolds. The $\pm$ entries correspond to a global $\pm$ sign action whereas $\frac{1}{2}-$ entries correspond to $x^i\mapsto-x^i+\frac{1}{2}$.}
\label{tab:discreteaction}
\end{table}}

{\renewcommand{\arraystretch}{1.3}
\begin{table}[h]
\begin{center}
\begin{tabular}{ |c||c|c|c|c|c|c|c|c| } 
\hline
   & $c_1$ & $c_2$ & $c_5$ & $c_6$ & $d_1$ & $d_3$ & $d_5$ & $d_7$ \\
 \hline
 \hline
 $I$ & $\frac{1}{2}$ & $\frac{1}{2}$ & $\frac{1}{2}$ & $\frac{1}{2}$ & $0$ & $\frac{1}{2}$ & $\frac{1}{2}$ & $\frac{1}{2}$ \\
 \hline
$II$ & $\frac{1}{2}$ & $0$ & $\frac{1}{2}$ & $0$ & $0$ & $\frac{1}{2}$ & $\frac{1}{2}$ & $\frac{1}{2}$ \\ 
\hline
$III$ & $\frac{1}{2}$ & $\frac{1}{2}$ & $\frac{1}{2}$ & $0$ & $0$ & $\frac{1}{2}$ & $\frac{1}{2}$ & $0$ \\ 
\hline
$IV$ & $\frac{1}{2}$ & $0$ & $\frac{1}{2}$ & $0$ & $0$ & $\frac{1}{2}$ & $\frac{1}{2}$ & $0$ \\
\hline
\end{tabular}
\end{center}
\caption{Coefficients for different orbifold examples (see Table~\ref{tab:discreteaction}).}
\label{tab:exampleslist}
\end{table}}

As explained in \cite{Braun:2018joh}, all these orbifolds admit a GCS realization pulling them apart along the coordinate $x^3=t$. In this case, the external circle corresponds to the coordinate $x^4=\theta$.\footnote{Alternative choices for the coordinates $(t,\theta)$ are possible depending on the orbifold of Table~\ref{tab:exampleslist} and we will comment on them later.}

From Table~\ref{tab:dictionary} we know that the chiral algebra corresponding to each coordinate $x^i$ of the orbifold is just a free algebra $(\psi_i,j_i)$. This can be used to provide a free field realization of the diamond of algebras associated to the GCS decomposition of the orbifolds. The top algebra $\text{Od}_3\oplus \text{Fr}^1_t\oplus \text{Fr}^1_\theta$ is given as follows. We have $\text{Fr}^1_t=\text{Fr}^1_3$ and $\text{Fr}^1_\theta=\text{Fr}^1_4$. The generators of $\text{Od}_3$ are given by
\begin{equation} \label{eq:Od3freefields}
\begin{split}
J^3_3 &= \ \normord{\psi_1\psi_2} +\normord{\psi_5\psi_6} +\normord{\psi_7\psi_8} \ , \\
A_3 &= \ \normord{\psi_1\psi_5\psi_8} +\normord{\psi_1\psi_6\psi_7} +\normord{\psi_2\psi_5\psi_7} -\normord{\psi_2\psi_6\psi_8} \ , \\
B_3 &= -\normord{\psi_1\psi_5\psi_7} +\normord{\psi_1\psi_6\psi_8} +\normord{\psi_2\psi_5\psi_8} +\normord{\psi_2\psi_6\psi_7} \ ,
\end{split}
\end{equation}
with the Virasoro generators $(T_3,G_3)$ given by the standard expressions \eqref{eq:virasorostandard} combining the coordinates 1, 2, 5, 6, 7 and 8. The realizations of the lateral tip algebras $\SVseven\oplus\text{Fr}^1$ and $\text{Od}_4$ are obtained directly from \eqref{eq:g2frsubod3fr2} and \eqref{eq:od4subod3fr2} respectively using the realization \eqref{eq:Od3freefields}. The bottom $\SVeight$ algebra is obtained either through \eqref{eq:spin7subg2fr} or \eqref{eq:spin7subod4} and it can be checked that the operator $-X_8$ matches the geometric expectation from \eqref{eq:spin7fourform}.

The composition of four T-duality automorphisms in any of the directions of \eqref{eq:listofTdualities} provides an automorphism that leaves invariant the diamond realization we have obtained. Each of the 14 different automorphisms therefore reduces to one of the possibilities we described in Table~\ref{tab:automorphisms}.

\begin{itemize}
    \item When no T-dualities are applied to $t$ or $\theta$, the automorphism corresponds to the identity $\textbf{A}_0$. This is the case for $\lbrace (1, 2, 5, 6), (1, 2, 7, 8), (5, 6, 7, 8) \rbrace$.
    \item When T-dualities are applied to both $t$ and $\theta$, the automorphism corresponds to $\textbf{A}_1$. This is the case for $\lbrace (1, 2, 3, 4), (3, 4, 5, 6), (3, 4, 7, 8)\rbrace$.
    \item When T-duality is applied to $t$ but not to $\theta$, the automorphism corresponds to $\textbf{A}_2$. This is the case for $\lbrace (1, 3, 5, 7), (1, 3, 6, 8), (2, 3, 5, 8), (2, 3, 6, 7)\rbrace$.
    \item When T-duality is applied to $\theta$ but not to $t$, the automorphism corresponds to $\textbf{A}_3$. This is the case for $\lbrace (1, 4, 5, 8), (1, 4, 6, 7), (2, 4, 5, 7), (2, 4, 6, 8)\rbrace$.
\end{itemize}

\noindent This shows that all our candidates for mirror automorphisms explicitly appear in these four orbifold examples. It is moreover satisfying to examine the geometric implementation of these maps as in \cite{Braun:2017ryx, Braun:2017csz}, as they agree with taking localized mirrors of the components of the GCS decomposition as suggested by Table~\ref{tab:automorphisms}. We do this presently.

\bigskip

Consider the four different $\textbf{A}_3$ mirror maps. In the ACyl G$_2$ end of the construction the external circle $\text{S}^1_\theta$ is T-dualized and the remaining $\mathbb{T}^3$ fibre is in all cases calibrated by the associative 3-form $\varphi.$\footnote{Let us illustrate this for the case where we apply T-dualities in the $(1, 4, 5, 8)$ directions. The coordinate $x^4$ corresponds to the external circle, so the $\mathbb{T}^3$ fibre in the ACyl G$_2$ end is given by the coordinates $x^1$, $x^5$ and $x^8$. It can be checked that the associative form $\varphi$ for this GCS realization has a $\dd x^{158}$ term. This means the restriction of $\varphi$ to the fibre is the volume form of the $\mathbb{T}^3$ and the fibre is calibrated by $\varphi$.} As explained earlier, this corresponds to a $\mathcal{T}^3$ mirror map in the G$_2$ manifold, which manifests in the algebra as a \textbf{GK} automorphism. In the neck region, the $\mathbb{T}^3$ fibre within the CY$_3$ is always calibrated by $\Re\Omega_3$ and it is therefore special Lagrangian. This means we are performing a mirror symmetry in the CY$_3$, which corresponds to an automorphism \textbf{M} in the chiral algebra. Finally, for the ACyl CY$_4$ end the whole $\mathbb{T}^4$ fibre is found to be calibrated by $\Re\Omega_4$, thus we have a mirror symmetry on the CY$_4$ and an \textbf{M} automorphism in the algebra.

Now let us explore the four automorphisms which reduce to $\textbf{A}_2$. Here in the ACyl G$_2$ end the $\mathbb{T}^4$ fibre is always calibrated by the coassociative 4-form $*\varphi$. This corresponds to a $\mathcal{T}^4$ mirror map in the G$_2$ manifold, which reduces to the identity in the chiral algebra. The neck region and the ACyl CY$_4$ end are similar to the $\textbf{A}_3$ case: we have T-duality in the $t$ direction and a $\mathbb{T}^3$ fibre calibrated by $\Im\Omega_3$ in the CY$_3$, whereas the whole $\mathbb{T}^4$ fibre is calibrated by $\Re\Omega_4$ in the CY$_4$. This means we expect mirror symmetry on the Calabi-Yau manifolds, which produces \textbf{M} automorphisms.

A general proposal to construct GCS mirror manifolds was given in \cite{Braun:2019lnn}: the idea is to apply mirror maps to the open ends of the construction and glue the manifolds back together. The mirror of the ACyl G$_2$ is obtained via a $\mathcal{T}^3$ map, and for the orbifolds presented this is precisely what $\textbf{A}_3$ describes. Our discussion above suggests the existence of an alternative mirror construction, based on the $\textbf{A}_2$ automorphism, where the map employed to obtain the ACyl G$_2$ mirror is $\mathcal{T}^4$. We discuss this further in Section~\ref{sec:NewMirrors}.

\bigskip

We now turn our attention to the three maps producing the $\textbf{A}_1$ automorphism. In the ACyl G$_2$ end we dualize the external circle and an associative $\mathbb{T}^3$ fibration, so we have a $\mathcal{T}^3$ mirror map and a \textbf{GK} automorphism. On the neck CY$_3$ we have a $\mathbb{T}^2$ fibration which is calibrated by the Hermitian form $\omega_3$. This means that this fibration is just a complex submanifold and is not supersymmetric, therefore these T-dualities do not correspond to a mirror symmetry on the CY$_3$ and the associated chiral algebra automorphism is just the identity. For the ACyl CY$_4$ end, the $\mathbb{T}^4$ fibration is calibrated by $\frac{1}{2}\omega_4\wedge\omega_4$ so again we find a complex submanifold and not a supersymmetric fibration, resulting in an identity automorphism.

Finally, let us study the three maps corresponding to $\textbf{A}_0$. The $\mathbb{T}^4$ fibre in the ACyl G$_2$ end turns out to be a coassociative fibration for the three maps. This means that even though we see an identity automorphism in the algebra, there is a non-trivial $\mathcal{T}^4$ mirror map acting on the G$_2$ manifold. The $\mathbb{T}^4$ fibre is calibrated by $\frac{1}{2}\omega_3\wedge\omega_3$ in the neck CY$_3$ and by $\frac{1}{2}\omega_4\wedge\omega_4$ in the ACyl CY$_4$, therefore it corresponds to complex submanifolds and the associated automorphisms are the identity in both cases.

Once again the geometric description is consistent with the automorphisms of Table~\ref{tab:automorphisms}. Moreover, the interpretation of these mirror maps is clear in these examples: we construct the mirror orbifold by applying a mirror map to the ACyl G$_2$ end of the construction whereas no mirror map is applied to the ACyl CY$_4$. When the $\mathcal{T}^3$ mirror map is applied we obtain an $\textbf{A}_1$ automorphism in the chiral algebra, whereas when the $\mathcal{T}^4$ map is applied the automorphism $\textbf{A}_0$ is obtained. We return to these in Section~\ref{sec:NewMirrors}.

\bigskip

Our choice of pulling the Joyce orbifolds along the $t=x^3$ direction crucially influenced the previous discussion, yet it is somewhat arbitrary. Some Joyce orbifolds admit more than one GCS decomposition. Let us briefly consider what changes if we stretch the orbifold along the $x^6$ direction. This can be done for the orbifold $I$, with the external circle given by the coordinate $x^8$.\footnote{Note that there are yet more possible GCS structures: orbifolds $I$ and $II$ can be pulled apart along the coordinate $x^7$, with the external circle in the coordinate $x^8$ and orbifolds $I$ and $III$ can be pulled along $x^2$ with a circle in $x^4$.} A realization of the diamond of algebras can be obtained and upon studying the action of T-dualities on it we find:

\begin{itemize}
    \item $\textbf{A}_0$ is obtained from $\lbrace (1, 2, 3, 4), (1, 3, 5, 7), (2, 4, 5, 7) \rbrace$.
    \item $\textbf{A}_1$ is obtained from $\lbrace (1, 3, 6, 8), (2, 4, 6, 8), (5, 6, 7, 8)\rbrace$.
    \item$\textbf{A}_2$ is obtained from $\lbrace (1, 2, 5, 6), (1, 4, 6, 7), (2, 3, 6, 7),  (3, 4, 5, 6)\rbrace$.
    \item$\textbf{A}_3$ is obtained from $\lbrace (1, 2, 7, 8), (1, 4, 5, 8), (2, 3, 5, 8), (3, 4, 7, 8)\rbrace$.
\end{itemize}

Note that most of the T-duality combinations are now assigned a different $\textbf{A}_i$. This illustrates that all automorphisms $\textbf{A}_i$ should equally be considered as mirror maps, since they may be exchanged into each other when more than one GCS decomposition is available.

To conclude, we mention for completeness the existence of combinations of T-dualities not included in \eqref{eq:listofTdualities} preserving the Cayley 4-form in these examples. The only possibilities are the trivial map, associated to $\textbf{A}_0$, and performing T-dualities along the eight coordinates, associated to $\textbf{A}_1$. These together with the maps \eqref{eq:listofTdualities} form a $\mathbb{Z}_2^4$ group under composition.

\subsubsection{New mirror maps} \label{sec:NewMirrors}

It is satisfying that T-dualities in Joyce orbifolds lead to recognizable mirror maps being applied to components of the GCS decomposition. 
Note that the interpretation of the diamond automorphisms $\mathbf{A}_i$ remains valid for more general GCS manifolds which do not present a description as a Joyce orbifold.
Extrapolating, this suggests that the mirror maps exist even in the general GCS setting.

We explained above how the GCS mirror construction proposed in \cite{Braun:2019lnn} corresponds to an $\textbf{A}_3$ automorphism in the associated diamond. We propose new methods to obtain mirrors of GCS manifolds based on the $\textbf{A}_2$, $\textbf{A}_1$ and $\textbf{A}_0$ automorphisms.

We begin with the $\textbf{A}_2$ construction. Consider a GCS manifold with a 4-tori fibration which is supersymmetric all over the manifold and which does not involve the external circle in the ACyl G$_2$ end. We can separate the two ends of the construction and, by dualizing the supersymmetric fibre, take a mirror map of type $\mathcal{T}^4$ in the ACyl G$_2$ and a mirror map in the ACyl CY$_4$. A mirror GCS manifold is obtained by gluing the ends back together after the mirror maps are applied.

The construction of mirrors based on the $\textbf{A}_1$ and $\textbf{A}_0$ automorphisms is different. Consider in this case a GCS manifold with a 4-tori fibration which is supersymmetric for the ACyl G$_2$ manifold but not for the ACyl CY$_4$. We can then construct a mirror GCS manifold by separating the two ends of the construction, taking a mirror map in the ACyl G$_2$ and four T-dualities in the ACyl CY$_4$, and gluing the ends back together. When the fibration includes the external circle in the ACyl G$_2$ end, the mirror map is of type $\mathcal{T}^3$ and the construction corresponds to an automorphism $\textbf{A}_1$. When the external circle is not included, the mirror map is $\mathcal{T}^4$ and the construction corresponds to $\textbf{A}_0$.

Note that in the $\textbf{A}_2$ construction we apply mirror maps to both ends of the GCS manifold, whereas for $\textbf{A}_1$ and $\textbf{A}_0$ this only occurs for the ACyl G$_2$ manifold. A similar phenomenon was observed in \cite{Braun:2017csz} for the mirror constructions of TCS manifolds associated to $\mathcal{T}^4$ and $\mathcal{T}^3$.
It is natural to ask if we could construct mirror GCS manifolds by applying a mirror map only to the ACyl CY$_4$ end of the construction and not to the ACyl G$_2$. Such a map would have to be associated to an $\textbf{A}_2$ automorphism with a non-coassociative fibre in the ACyl G$_2$. We do not find such a map amongst the examples we studied.

There is another piece of evidence for our GCS mirror symmetry proposal. One of the main arguments provided in \cite{Braun:2019lnn} in support of their GCS mirror symmetry construction ($\mathbf{A}_3$ in our notation) was the invariance of the dimension of the $\sigma$-model moduli space under the mirror maps, at least under some simplifying assumptions,
\begin{equation} \label{eq:modulispacedim}
    b^2(\mathcal{M})+b^4_-(\mathcal{M})+1=b^2(\mathcal{M^\vee})+b^4_-(\mathcal{M^\vee})+1 \ ,
\end{equation}
where $b^i$ denotes the Betti number of dimension $i$, the subscript $-$ indicates a restriction to anti-self dual forms, $\mathcal{M}$ denotes the GCS manifold and $\mathcal{M^\vee}$, its mirror. The proof of \eqref{eq:modulispacedim} detailed in {Section~4.2 of} \cite{Braun:2019lnn} {works as follows: as a first step the Betti numbers appearing in the LHS of \eqref{eq:modulispacedim} are rewritten in terms of cohomology groups of the open ends and the neck region by a Mayer-Vietoris argument. One then studies the effect on these cohomologies of applying mirror maps on both open ends of $\mathcal{M}$, eventually reaching the RHS of \eqref{eq:modulispacedim}, as we now describe.

Consider the ACyl CY$_4$ end. By gluing together two copies of this end one obtains a compact CY$_4$. The mirror map of the ACyl CY$_4$ extends to a mirror map of the compact CY$_4$ which changes the Hodge numbers in the usual way: $h^{i,j}(\text{CY}_4)=h^{4-i,4-j}(\text{CY}_4^\vee)$. These numbers are related to those of the ACyl CY$_4$ and the neck region by another Mayer-Vietoris sequence, and one deduces from here restrictions on how the cohomologies change under mirror symmetry. It turns out the LHS of \eqref{eq:modulispacedim} is invariant under these changes.

An analogous argument for the ACyl G$_2$ end using the fact that G$_2$ mirror symmetry preserves the combination $b^2+b^3$ shows that the LHS of \eqref{eq:modulispacedim} is also preserved by a mirror symmetry on the ACyl G$_2$ end, showing that the equality \eqref{eq:modulispacedim} holds. This proof works for both $\mathcal{T}^3$ and $\mathcal{T}^4$ mirror maps, so in particular it also holds for the $\mathbf{A}_2$ construction.}

{Moreover, the invariance of the LHS of \eqref{eq:modulispacedim} under mirror symmetry on the ACyl CY$_4$ end is independent from the ACyl G$_2$ end and vice versa. This means that if a mirror map is applied to just one of the open ends, \eqref{eq:modulispacedim} remains valid. This is the case of the proposed $\mathbf{A}_0$ and $\mathbf{A}_1$ constructions, where the mirror map is applied only to the ACyl G$_2$ manifold, leaving the cohomologies of the ACyl CY$_4$ intact. Therefore, the dimension of the moduli space is preserved by all our proposed mirror constructions. Note also that \eqref{eq:modulispacedim} would still hold in a construction where a mirror map is applied only to the ACyl CY$_4$ end.}

\section{Are Connected Sums Generic?} \label{sec:Num}

In this section we compare the algebra at the bottom of the diamond diagram---either Figure~\ref{fig:TCS}(b) for (E)TCS or Figure~\ref{fig:GCS}(b) for GCS---and the intersection of the algebras on the lateral tips. By construction the bottom algebra is contained in the intersection algebra reflecting the fact that the manifold has holonomy G$_2$ or Spin(7) respectively.

Suppose this inclusion was strict, say for the GCS diagram for definiteness. This would mean there are fields in the intersection algebra that do not appear in the bottom algebra, leading to additional chiral symmetries. As a result, we would conclude that GCS manifolds have a set of symmetries larger than a generic manifold of holonomy Spin(7), whose chiral algebra is simply $\SVeight$. We postulate however that the opposite is true, i.e.\ that the intersection algebra perfectly agrees with the bottom algebra of the diamond both for (E)TCS and GCS. This would mean that, at least from a chiral algebra viewpoint, these constructions are representative of generic manifolds with holonomy G$_2$ or Spin(7).

The vacuum module character of a chiral algebra is defined as
\begin{equation}
    \chi=\tr\left(q^{L_0-\frac{c}{24}}\right),
\end{equation}
where the trace is taken over the vacuum module. The number of independent fields at level $h$ of the chiral algebra can be read off from the expansion of this character in powers of $q=e^{2\pi i \tau}$, as the coefficient of $q^{h-c/24}$.

An analytic proof of our proposal would require a good grasp of the vacuum module characters not only of the $\SVseven$ and $\SVeight$ algebras but also of the intersections involved. Unfortunately the latter seem very challenging to obtain, and only the character of $\SVeight$ is known analytically \cite{Benjamin:2014kna}.

We therefore rely on numerical checks to support the proposal. For each chiral subalgebra appearing in the diamond, the corresponding character expansion is obtained by listing and counting the fields in the subalgebra level by level. For the intersection of the lateral tip subalgebras, we search at each level for all linear combinations of fields which are contained in both algebras. The number of linearly independent combinations at each level provides the coefficients in the character expansion of the intersection algebras.

The numerical manipulations become more involved as we look into higher levels due to the dramatic increase in the number of null fields. Indeed the top algebras of both diagrams have singular fields whose descendants are null and must be quotiented out at each subsequent levels. When listing the fields in a subalgebra or an intersection, one has to ensure that this quotient is taken into account.

Once this is achieved we obtain vacuum character expansions of the different algebras and these can be compared to test the proposal. We verify the agreement level by level in the various vacuum modules.

\subsection{The (E)TCS case}

The precise embedding of $(\text{Od}_3\oplus \text{Fr}^1)_-$ inside $\text{Od}_2\oplus \text{Fr}^3$ depends on the gluing angle of the ETCS construction we are considering. We have a circle worth of embeddings where any two of them can be mapped to each other by an automorphism leaving the underlying $\SVseven$ invariant. We first perform the computation for the TCS embedding corresponding to a gluing angle of $\vartheta=\pi/2$.

As we mentioned earlier, an analytic expression for the character of the $\SVseven$ algebra is unfortunately not available in the literature so far. We have however managed to compute it numerically up to level 9, and we obtained
\begin{align} \label{eq:svG2character}
    \chi(\SVseven)=&q^{-\frac{7}{16}}\bigg(1+2q^{3/2}+3 q^2+3 q^{5/2}+4 q^3+8 q^{7/2}+12 q^4+14 q^{9/2}+18 q^5+29 q^{11/2}\nonumber \\
    &+42 q^{6}+51 q^{13/2}+66 q^{7}+96 q^{15/2}+129 q^{8}+160 q^{17/2}+207 q^{9}+
    \ldots
    \bigg) \ .
\end{align}
Each coefficient in the expansion \eqref{eq:svG2character} corresponds to the number of independent fields of $\SVseven$ at the level given by the corresponding power of $q$. We have been able to verify numerically up to level 5 that these agree with the number of independent fields of $(\text{Od}_3\oplus \text{Fr})_+\cap(\text{Od}_3\oplus \text{Fr})_-$, thus giving evidence for our proposal in this case.

For the case of arbitrary gluing angle $\vartheta\in (0,\pi)$, the nontrivial trigonometric functions involved make the computation harder and we have only managed to verify the equality numerically up to level 3.

\subsection{The GCS case}

Now consider the GCS diagram,  Figure~\ref{fig:GCS}(b). In this case we want to check that the intersection of the $\text{Od}_4$ and $\SVseven \oplus \text{Fr}^1$ subalgebras of $\text{Od}_3\oplus \text{Fr}^2$ is precisely the $\SVeight$ subalgebra.

The vacuum character of $\SVeight$ was recently computed in \cite{Benjamin:2014kna} and it is given by
\begin{align}
    \chi(\SVeight)=q^{-\frac{1}{2}}{\mathcal P}(\tau)\Bigg(1-\sum_{k=0}^\infty \Big(q^{\frac{15}{2}k^2+4k+\frac{1}{2}} +\frac{q^{\frac{15}{2}k^2+2k+\frac{1}{2}}}{1+q^\frac{6k+1}{2}} -\frac{q^{\frac{15}{2}k^2+7k+2}}{1+q^\frac{6k+3}{2}}- \nonumber
    \\
    -q^{\frac{15}{2}k^2+14k+\frac{13}{2}} +\frac{q^{\frac{15}{2}k^2+14k+\frac{11}{2}}}{1+q^\frac{6k+3}{2}} -\frac{q^{\frac{15}{2}k^2+19k+11}}{1+q^\frac{6k+5}{2}} \Big)\Bigg) \ ,
\end{align}
where
\begin{equation}
    {\mathcal P}(\tau)=\prod_{k=1}^\infty\left( \frac{1+q^{k-1/2}}{1-q^k} \right)^2 \ .
\end{equation}
Expanding in powers of $q$, the first few terms are
\begin{align} \label{eq:svSpin7character}
    \chi(\SVeight)=q^{-\frac{1}{2}}&\bigg(1+q^{3/2}+2 q^2+2 q^{5/2}+2 q^3+4 q^{7/2}+7 q^4 \nonumber \\
    &\qquad\qquad\qquad+8 q^{9/2}+9 q^5+14 q^{11/2}+21 q^{6}
    +\ldots
    \bigg) \ .
\end{align}
We have checked that the number of independent fields in the intersection algebra matches this expansion precisely up to level 6. Since we already know that the $\SVeight$ algebra is contained in the intersection, this is enough to show the equality between these algebras up to level 6 and provides a check to our proposal.

\section{Conclusion}

In this paper we have explored the relationship between the geometry of connected sum manifolds $\mathcal{M}$ of holonomies G$_2$ and Spin(7), and the chiral algebra of the associated $\sigma$-model. Starting from the geometric description of $\mathcal{M}$ in terms of open patches, we have argued for a diamond of algebra inclusions in the worldsheet theory.
We have shown the validity of the diamond for Extra Twisted Connected Sum (ETCS) G$_2$-manifolds, and Generalized Connected Sum (GCS) Spin(7)-manifolds. We have checked numerically the agreement, at leading orders, between the Shatashvili-Vafa algebra at the bottom of the diamond and the intersection of the algebras at the lateral tips of the diamond, which suggests that these constructions provide generic special holonomy manifolds.

Additionally, we have {described all}
the {possible} automorphisms fixing the GCS diamond and we have interpreted them in terms of GCS mirror symmetry maps. {In the case of four different Joyce orbifolds, we have shown that every mirror map coming from T-dualities corresponds to one of these GCS mirror maps}.
This has lead us to propose new constructions of GCS mirror manifolds.

Our results set the ground for different future directions. A natural next step would be to study the diamond of algebras for other manifolds constructed by gluing two building blocks. Some Calabi-Yau manifolds can be obtained by this procedure; an example is the Schoen Calabi-Yau 3-fold \cite{Schoen1988}, see \cite{Braun:2017uku}. Further examples of Calabi-Yau 4-folds and G$_2$ manifolds built from gluing two ACyl copies can be found in \cite{Braun:2018joh}.

In our analysis of mirror maps for Spin(7) Joyce orbifolds we did not find any
that was acting as mirror symmetry on the ACyl CY$_4$ end while the ACyl G$_2$ end was dualized along a non-coassociative fibre. The existence of such a map is an intriguing possibility and one could try to look for an explicit realization in GCS manifolds beyond Joyce orbifolds such as the ones described in \cite{Braun:2018joh}. We know it would have to correspond to an $\textbf{A}_2$ automorphism.

It would also be interesting to perform a thorough study of the automorphisms available for the ETCS diamond in the same spirit as Section~\ref{sec:Spin7Aut}. We could then try to give an interpretation of these as mirror maps for ETCS manifolds and look for realizations in explicit examples. In particular for the TCS case we should recover the maps $\mathcal{T}^3$ and $\mathcal{T}^4$ of \cite{Acharya:1997rh}, \cite{Gaberdiel:2004vx}.

The general interpretation of our results given in Section~\ref{sec:Idea} was in terms of worldsheet symmetries of local patches of the $\sigma$-model target space. Although this was sufficient for us, it is hard to miss the similarities with the so-called chiral de Rham complex, which is a sheaf of vertex operator algebras \cite{Malikov:1998dw}. In the context of $\sigma$-models, it is thought to describe localized fluctuations in the target space geometry, much in the spirit we have advocated \cite{Witten:2005px}. It would be interesting to understand precisely this connection.

Another tempting direction is to attempt to find modular invariant partition functions for exceptional holonomy manifolds by capitalizing on the variety of Calabi-Yau vacua under worldsheet control.
It would be particularly interesting to search for rational Od$_n$ CFTs which could be used as the ``theory on the neck region'' either for G$_2$ ($n=2$) or Spin(7) ($n=3$) backgrounds. They would also have to feature the corresponding diamond of subsymmetries we have presented, which constrains the possible choices.

\section*{Acknowledgements}

We are grateful to Xenia de la Ossa for comments on a preliminary version of this paper. We thank Sebastian Goette for suggesting to consider ETCS, and Matthias Gaberdiel for general guidance and for asking the question leading to Section~\ref{sec:Num}. We also acknowledge discussions with Andreas Braun, Christopher Beem and Suvajit Majumder. MAF is supported by an SNF grant and by
the NCCR SwissMAP, that is also funded by the Swiss National Science Foundation.
MG is supported by a scholarship from the Mathematical Institute, University of Oxford, as well as a fellowship from ``la Caixa” Foundation (ID 100010434) with fellowship code LCF/BQ/EU17/11590062.

\appendix

\section{The Od$_4$ algebra} \label{app:Od4}

Here we present the nontrivial OPE relations we worked out for the Od$_4$ algebra.

\begin{align*}
T_4(z)T_4(w)&=\frac{6}{(z-w)^4}+\frac{2
   T(w)}{(z-w)^2}+\frac{\partial T(w)}{z-w}+\cdots,\\
T_4(z)G_4(w)&=\frac{3 G_4(w)}{2 (z-w)^2}+\frac{\partial G_4(w)}{z-w}+\cdots,\\
T_4(z)G^3_4(w)&=\frac{3 G^3_4(w)}{2 (z-w)^2}+\frac{\partial G^3_4(w)}{z-w}+\cdots,\\
T_4(z)J^3_4(w)&=\frac{J^3_4(w)}{(z-w)^2}+\frac{\partial J^3_4(w)}{z-w}+\cdots,\\
G_4(z)G_4(w)&=\frac{8 }{(z-w)^3}+\frac{2 T(w)}{z-w}+\cdots,\\
G_4(z)G^3_4(w)&=\frac{2 J^3_4(w)}{(z-w)^2}+\frac{\partial J^3_4(w)}{z-w}+\cdots,\\
G_4(z)J^3_4(w)&=\frac{G^3_4(w)}{z-w}+\cdots,\\
G^3_4(z)G^3_4(w)&=\frac{8 }{(z-w)^3}+\frac{2 T(w)}{z-w}+\cdots,\\
G^3_4(z)J^3_4(w)&=-\frac{G_4(w)}{z-w}+\cdots,
\end{align*}

\begin{align*}
J^3_4(z)J^3_4(w)&=-\frac{4 }{(z-w)^2}+\cdots,\\
T_4(z)A_4(w)&=\frac{2 A_4(w)}{(z-w)^2}+\frac{\partial A_4(w)}{z-w}+\cdots,\\
T_4(z)B_4(w)&=\frac{2 B_4(w)}{(z-w)^2}+\frac{\partial B_4(w)}{z-w}+\cdots,\\
T_4(z)C_4(w)&=\frac{5 C_4(w)}{2 (z-w)^2}+\frac{\partial C_4(w)}{z-w}+\cdots,\\
T_4(z)D_4(w)&=\frac{5 D_4(w)}{2 (z-w)^2}+\frac{\partial D_4(w)}{z-w}+\cdots,\\
G_4(z)A_4(w)&=\frac{C_4(w)}{z-w}+\cdots,\\
G_4(z)B_4(w)&=\frac{D_4(w)}{z-w}+\cdots,\\
G_4(z)C_4(w)&=\frac{4 A_4(w)}{(z-w)^2}+\frac{\partial A_4(w)}{z-w}+\cdots,\\
G_4(z)D_4(w)&=\frac{4 B_4(w)}{(z-w)^2}+\frac{\partial B_4(w)}{z-w}+\cdots,\\
G^3_4(z)A_4(w)&=-\frac{D_4(w)}{z-w}+\cdots,\\
G^3_4(z)B_4(w)&=\frac{C_4(w)}{z-w}+\cdots,\\
G^3_4(z)C_4(w)&=\frac{4 B_4(w)}{(z-w)^2}+\frac{\partial B_4(w)}{z-w}+\cdots,\\
G^3_4(z)D_4(w)&=-\frac{4 A_4(w)}{(z-w)^2}-\frac{\partial A_4(w)}{z-w}+\cdots,\\
J^3_4(z)A_4(w)&=-\frac{4 B_4(w)}{z-w}+\cdots,\\
J^3_4(z)B_4(w)&=\frac{4 A_4(w)}{z-w}+\cdots,\\
J^3_4(z)C_4(w)&=-\frac{3 D_4(w)}{z-w}+\cdots,\\
J^3_4(z)D_4(w)&=\frac{3 C_4(w)}{z-w}+\cdots,\\
A_4(z)A_4(w)&=\frac{8}{(z-w)^4}-\frac{4}{(z-w)^2}\normord{J^3_4J^3_4}(w)-\frac{4}{z-w}\normord{\partial J^3_4J^3_4}(w)+\cdots,\\
A_4(z)B_4(w)&=\frac{8}{(z-w)^3}J^3_4(w)+\frac{4}{(z-w)^2}\partial J^3_4(w)+\\
&+\frac{4/3}{z-w}\left(-\normord{J^3_4J^3_4J^3_4}+\partial\partial J^3_4\right)(w)+\cdots,\\
A_4(z)C_4(w)&=-\frac{4}{(z-w)^3}G_4(w)-\frac{4}{(z-w)^2}\left(\normord{G^3_4 J^3_4}+\partial G_4\right)(w)+\\
&+\frac{1}{z-w}\left(2\normord{G_4J^3_4J^3_4}-2\normord{G^3_4 \partial J^3_4}-4\normord{\partial G^3_4 J^3_4}-2\partial\partial G_4\right)(w)+\cdots,\\
A_4(z)D_4(w)&=\frac{4}{(z-w)^3}G^3_4(w)-\frac{4}{(z-w)^2}\left(\normord{G_4 J^3_4}-\partial G^3_4\right)(w)+\\
&+\frac{1}{z-w}\left(-2\normord{G^3_4J^3_4J^3_4}-2\normord{G_4 \partial J^3_4}-4\normord{\partial G_4 J^3_4}+2\partial\partial G^3_4\right)(w)+\cdots,
\end{align*}

\begin{align*}
B_4(z)B_4(w)&=\frac{8}{(z-w)^4}-\frac{4}{(z-w)^2}\normord{J^3_4J^3_4}(w)-\frac{4}{z-w}\normord{\partial J^3_4J^3_4}(w)+\cdots,\\
B_4(z)C_4(w)&=-\frac{4}{(z-w)^3}G^3_4(w)+\frac{4}{(z-w)^2}\left(\normord{G_4 J^3_4}-\partial G^3_4\right)(w)+\\
&+\frac{1}{z-w}\left(2\normord{G^3_4J^3_4J^3_4}+2\normord{G_4 \partial J^3_4}+4\normord{\partial G_4 J^3_4}-2\partial\partial G^3_4\right)(w)+\cdots,\\
B_4(z)D_4(w)&=-\frac{4}{(z-w)^3}G_4(w)-\frac{4}{(z-w)^2}\left(\normord{G^3_4 J^3_4}+\partial G_4\right)(w)+\\
&+\frac{1}{z-w}\left(2\normord{G_4J^3_4J^3_4}-2\normord{G^3_4 \partial J^3_4}-4\normord{\partial G^3_4 J^3_4}-2\partial\partial G_4\right)(w)+\cdots,\\
C_4(z)C_4(w)&=-\frac{32}{(z-w)^5}+\frac{8}{(z-w)^3}\left(\normord{J^3_4J^3_4}-T_4\right)(w)+\\
&+\frac{1}{(z-w)^2}\left(8 \normord{\partial J^3_4J^3_4}-4 \partial T_4\right)(w)+\\
&+\frac{1}{z-w}\big(-4\normord{G_4G^3_4J^3_4}-2\normord{G_4\partial G_4}-2\normord{G^3_4\partial G^3_4}+\\
&\hspace{1.5cm}+2\normord{\partial J^3_4\partial J^3_4}+4\normord{T_4J^3_4J^3_4}\big)(w)+\cdots,\\
C_4(z)D_4(w)&=-\frac{24}{(z-w)^4}J^3_4(w)-\frac{12}{(z-w)^3}\partial J^3_4(w)+\\
&+\frac{1}{(z-w)^2}\left(4\normord{G_4 G^3_4}+\frac{4}{3}\normord{J^3_4J^3_4J^3_4}-8 \normord{T_4J^3_4}-\frac{4}{3}\partial\partial J^3_4\right)(w)+\\
&+\frac{1}{z-w}\bigg(2\normord{\partial G_4 G^3_4}+2\normord{G_4\partial G^3_4}+2\normord{\partial J^3_4J^3_4J^3_4}-4 \normord{T_4\partial J^3_4}-\\
&\hspace{1.5cm}-4 \normord{\partial T_4J^3_4}+\frac{1}{3}\partial\partial\partial J^3_4\bigg)(w)+\cdots,\\
D_4(z)D_4(w)&=-\frac{32}{(z-w)^5}+\frac{8}{(z-w)^3}\left(\normord{J^3_4J^3_4}-T_4\right)(w)+\\
&+\frac{1}{(z-w)^2}\left(8 \normord{\partial J^3_4J^3_4}-4 \partial T_4\right)(w)+\\
&+\frac{1}{z-w}\big(-4\normord{G_4G^3_4J^3_4}-2\normord{G_4\partial G_4}-2\normord{G^3_4\partial G^3_4}+\\
&\hspace{1.5cm}+2\normord{\partial J^3_4\partial J^3_4}+4\normord{T_4J^3_4J^3_4}\big)(w)+\cdots.
\end{align*}


\normalem

\newcommand{\etalchar}[1]{$^{#1}$}

\end{document}